\documentclass[11pt]{article}
\usepackage{float}

\usepackage{bm, commath}
\usepackage{natbib}
\usepackage{caption}
\usepackage{graphicx}
\usepackage{subfig}
\usepackage{amsmath, amsfonts, amsthm}
\usepackage{float}
\usepackage{booktabs,siunitx}
\usepackage{url}
\usepackage{multirow}
\usepackage{amssymb}
\usepackage{blkarray}
\usepackage{bbm}
\usepackage{multicol}
\usepackage{authblk}

\usepackage{listings}
\usepackage{bbm}
\usepackage{textcomp}
\usepackage[margin=1in]{geometry}
\usepackage{authblk}
\usepackage[ruled]{algorithm2e}
\SetKwInput{KwParam}{Parameter}
\SetAlgoCaptionLayout{centerline}

\usepackage{sectsty}
\usepackage[onehalfspacing]{setspace}
\usepackage{tikz}
\usepackage{array}
\newcolumntype{P}[1]{>{\centering\arraybackslash}p{#1}}
\usetikzlibrary{shapes,decorations,arrows,calc,arrows.meta,fit,positioning}
\tikzset{
    -Latex,auto,node distance =1 cm and 1 cm,semithick,
    state/.style ={ellipse, draw, minimum width = 0.7 cm},
    point/.style = {circle, draw, inner sep=0.04cm,fill,node contents={}},
    bidirected/.style={Latex-Latex,dashed},
    el/.style = {inner sep=2pt, align=left, sloped}
}

\newcommand{\indep}{\rotatebox[origin=c]{90}{$\models$}}

\newtheorem{definition}{Definition}

\newtheorem*{assumption*}{Assumption}

\newtheorem{proposition}{Proposition}

\usepackage{mathtools}

\usepackage{xcolor}

\makeatletter
\renewcommand{\algocf@captiontext}[2]{#1\algocf@typo. \AlCapFnt{}#2} % text of caption
\def\@algocf@capt@plain{top}
\renewcommand{\algocf@makecaption}[2]{%
  \addtolength{\hsize}{\algomargin}%
  \sbox\@tempboxa{\algocf@captiontext{#1}{#2}}%
  \ifdim\wd\@tempboxa >\hsize%     % if caption is longer than a line
  \hskip .5\algomargin%
  \parbox[t]{\hsize}{\algocf@captiontext{#1}{#2}}% then caption is not centered
  \else%
  \global\@minipagefalse%
  \hbox to\hsize{\box\@tempboxa}% else caption is centered
  \fi%
  \addtolength{\hsize}{-\algomargin}%
}
\makeatother

\begin{document}

\sectionfont{\bfseries\large\sffamily}%

\subsectionfont{\bfseries\sffamily\normalsize}%

%\begin{center}
%\noindent
%{\sffamily\bfseries\LARGE 
%}%

%\noindent

%\end{center}

\title{Sensitivity Analysis for Attributable Effects in Case$^2$ Studies}

\author[1]{Kan Chen  }
\author[2]{Ting Ye }
\author[3]{Dylan S. Small \thanks{Email: {\tt dsmall@wharton.upenn.edu.}}}

\affil[1]{Department of Biostatistics, Harvard University, Boston, MA, U.S.A.}
\affil[2]{Department of Biostatistics, University of Washington, Seattle, WA, U.S.A.}
\affil[3]{Department of Statistics and Data Science, The Wharton School, University of Pennsylvania, Philadelphia, PA, U.S.A.}

\date{}

\maketitle

\noindent

%\textsf{{\bf Abstract}:} \input{abstract}

\begin{abstract}
The case$^2$ study, also referred to as the case-case study design, is a valuable approach for conducting inference for treatment effects. Unlike traditional case-control studies, the case$^2$ design compares treatment in two types of cases with the same disease.  A key quantity of interest is the attributable effect, which is the number of cases of disease among treated units which are caused by the treatment.  Two key assumptions that are usually made for making inferences about the attributable effect in case$^2$ studies are 1.) treatment does not cause the second type of case, and 2.) the treatment does not alter an individual's case type. However, these assumptions are not realistic in many real-data applications. In this article, we present a sensitivity analysis framework to scrutinize the impact of deviations from these assumptions on obtained results. We also include sensitivity analyses related to the assumption of unmeasured confounding, recognizing the potential bias introduced by unobserved covariates. The proposed methodology is exemplified through an investigation into whether having violent behavior in the last year of life increases suicide risk via 1993 National Mortality Followback Survey dataset. 
\end{abstract}

\vspace{0.3 cm}

\section{Introduction}
\label{sec: intro}

\subsection{Background: Suicide Study,  Case$^2$-Studies, Attributable Effect, and Selection Bias}

Suicide is a critical public health issue that necessitates a deep understanding of its various risk factors to prevent tragic outcomes effectively \citep{curtin2016increase,turecki2019suicide}. Among the myriad factors, violent behavior could be a particularly alarming indicator \citep{conner2001violence}. To study the effect of violent behavior on suicide risk, we adopt a case$^2$ study to delve deeper into this relationship.

Case$^2$ studies, also known as case-case studies, represent an innovative approach to harness datasets exclusively comprised of cases for investigating treatment effects that may potentially influence a subset of these events. Such datasets, devoid of non-case observations, present a distinctive structure where every recorded entry falls within the category of interest, such as disease occurrences, specific events, or rare incidents \citep{mccarthy1999case,campylobacter2002ciprofloxacin,pogreba2020methodology}. For example, the Fatal Accident Reporting System (FARS), commonly used in traffic safety studies, contains detailed records of fatal vehicle accidents in the United States but lacks data on non-accidents or trips that did not result in a crash. 

%\textcolor{blue}{The key distinction between case-case and case-control studies is that case-case studies compare different subtypes within cases, while case-control studies compare cases to non-cases. A common confusion is why we cannot treat one subtype as a non-case and perform a case-control study instead of a case-case study. This is because both subtypes represent cases of the condition or event, albeit with different characteristics or causes. For instance, in a case-case study comparing two types of Salmonella infections (e.g., Salmonella enteritidis vs. other Salmonella serovars) \citep{mccarthy1999case}, both subtypes involve individuals with the infection (i.e., both are cases). One cannot classify those with ``other Salmonella serovars'' as non-cases because they still have the infection, albeit with a different strain. A true non-case would be someone without any Salmonella infection. Misclassifying one infected group as non-cases would lead to flawed comparisons, as both groups experienced the outcome, albeit from different strains.}

Although the absence of a reference group or non-cases presents a challenge for comparative analysis, opportunities to study the causal effects of treatments on cases remain viable. Rosenbaum's work advocated for using the attributable effect to analyze the causal effect of a specific treatment in case$^2$ studies \citep{rosenbaum2002overt,rosenbaum2005attributable} by comparing two types of cases, one which could be caused by the treatment and one which could not be caused by the treatment. The attributable effect is the number of cases caused by the treatment.      
Inferential procedures regarding the attributable effect align with standard statistical methodologies. This is because the attributable effect effectively summarizes a multitude of potential values within a high-dimensional parameter space, making it applicable within the framework of traditional statistical analysis \citep{rosenbaum2001effects}. Research on the attributable effect has been conducted in matched observational studies \citep{rosenbaum2001effects}, case-control studies \citep{rosenbaum2002attributing}, and case$^2$ studies \citep{rosenbaum2005attributable}. Additional discussions of attributable effects can be found in \citet{macmahon1970epidemiology,walter1978calculation}.

%Consider two types of cases: narrow case and marginal case \citep{small2013case} where the narrow case has a greater effect size than the marginal case. 

Consider two types of cases: narrow cases and marginal cases \citep{rosenbaum2005attributable,small2013case}. The treatment can cause narrow cases but cannot cause marginal cases. Broad cases include both narrow and marginal cases. In conducting inference for the attributable effect in case$^2$ studies, two crucial assumptions come into play. Firstly, the treatment must not induce the occurrence of marginal cases. Secondly, the treatment must not cause a change in an individual's case type, implying no selection bias (i.e., no bias results from selecting samples based on characteristics affected by the treatment). In other words, the treatment does not alter an individual’s case type such that it only affects the outcome by transitioning an individual from non-case status to case status ---it does not shift an individual from a narrow case to a marginal case, nor from a marginal case to a narrow case. If these two important assumption hold, then a case$^2$ study can be treated like a traditional case-control study. In this situation, the other subtype can be treated as the non-case. This article addresses performing statistical inference for the attributable effect in case$^2$ studies in the presence of selection bias, given that both assumptions are quite uncertain in specific contexts, such as our motivating example: the study of violent behavior, suicide risk, and accidental death. In the context of our data application, suicide death is the narrow case while accidental death is the marginal case. The first assumption implies that violent behavior will never lead to accidental death. The second assumption implies that violent behavior will not change a suicide death to an accidental death, nor vice versa.

\subsection{Motivating Example: Violent Behavior, Suicide Risk, and Accidental Death}

\citet{conner2001violence} examined the association between violent behavior in the last year of life and completed suicide, accounting for alcohol use disorders using data from the 1993 National Mortality Followback Survey (NMFS) \citep{nmfs1993}. They found that violent behavior is a distinguishing factor between suicide victims and accidental death victims, and that this distinction is not solely due to alcohol use disorders. This suggests that initiatives aimed at preventing violence may also reduce suicide risks. However, there are two potential issues with this study: 1.) violent behavior may elevate the risk of unintentional injuries that could lead to accidental deaths \citep{gray2014comparative} (violation of the first assumption); and 2.) the presence of violent behavior might complicate the classification of deaths, making it difficult to differentiate between accidental deaths and suicides, as illustrated by cases involving driver suicides \citep{ohberg1997driver}  (violation of the second assumption). Addressing these issues is crucial in the context of this study.

In our exploration of the connection between violent behavior and suicide, we employ the 1993 NMFS dataset as our primary source of information as well. 
This dataset covers individuals aged 15 and older who died in 1993, with death certificate sampling approved by 49 states, the District of Columbia, and New York City. The sampled 22,957 death certificates from the 1993 Current Mortality Sample (CMS). Some subgroups were oversampled including black decedents, female decedents, decedents under age 35 and decedents over age 99 from the CMS. Further details can be found at \url{https://www.cdc.gov/nchs/nvss/nmfs/nmfs_methods.htm}. By utilizing this dataset, we aim to analyze and understand the intricate relationships and patterns within the realms of violent behavior, suicide deaths, and accidental deaths. Specifically, we focus on the causal question: does having violent behavior in the last year of life affect suicide deaths in comparison to accidental deaths? 

%By utilizing this dataset, we aim to analyze and understand the intricate relationships and patterns within the realms of violence and suicide. Specifically, we investigate a case study via NMFS 1993 dataset: whether having violent behavior in the last year of life increases suicide risk in comparison to accidental deaths. 

%The 1993 NMFS offers a valuable glimpse into the patterns of mortality during that period, providing comprehensive data on various aspects of death in the United States. Conducted by the National Center for Health Statistics, this survey collected information through interviews with next-of-kin or other knowledgeable individuals about the deceased. The dataset includes details on causes of death, contributing factors, demographics, and other relevant information. Analyzing the data from the 1993 NMFS enables researchers and policymakers to gain insights into the epidemiology of mortality, identify trends, and understand the factors influencing health outcomes. 

In our study employing the case-case design with the 1993 NMFS, which solely contains death information without non-cases, we focus on the attributable effect of violent behavior on death outcomes. Given that violent behavior could lead to accidental deaths, it challenges the assumption that treatments won't affect the occurrence of such cases. Moreover, violent behavior might not only increase accidental deaths but also has the potential to convert what could be classified as accidental deaths into suicides, or conversely, motivate suicides that could be misconstrued as accidents. This interplay introduces a significant risk of selection bias. Consequently, it is essential to develop a sensitivity analysis framework that accounts for these assumptions. Such a framework is critical for accurately inferring the attributable effects in a case-case study design, ensuring that the analysis remains robust despite the complexities introduced by the nature of violent behavior and its impact on mortality categorizations.

\subsection{Our Contribution, Prior Work, and Outline}

%In this paper, we introduce an innovative sensitivity analysis framework tailored to address the potential violation of assumptions aforementioned in the inference for attributable effects within matched case$^2$ studies. Recognizing the crucial importance of mitigating bias in such analyses, our framework aims to enhance the robustness and reliability of estimates across diverse settings. Specifically applied to investigate the attributable effect of violent behavior on the risk of suicide, our methodology systematically examines potential selection bias impacts. This approach provides a nuanced understanding of the relationships involved, ensuring that the results are robust and accurately reflect the influence of violent behavior on suicide risk within the study's context. Furthermore, our sensitivity analysis framework incorporates analysis for hidden bias as well, offering a comprehensive examination of potential sources of bias. 

In this paper, we introduce a novel sensitivity analysis framework designed to address potential violations of the assumptions underlying the inference of attributable effects in matched case$^2$ studies. Specifically, in addition to the sensitivity parameter for unmeasured confounding introduced by \citet{rosenbaum2002overt}, we incorporate two additional sensitivity parameters to account for the violation of the two key assumptions mentioned in the last paragraph of Section 1.1 and provide a bounding analysis. We apply this framework to a study examining the impact of violent behavior on the risk of suicide using 1993 NMFS dataset.

In prior work on this question, \citet{conner2001violence} analyzed 1993 NMFS data on 753 suicides and 2,115 accidents (ages 20-64). Using the CAGE questionnaire, they assessed violent behavior and alcohol misuse history. Multiple logistic regression revealed violent behavior as a distinguishing factor between suicide and accident victims, irrespective of alcohol use disorders. This underscores the role of violence prevention initiatives in mitigating overall suicide risk. However, their work did not take into consideration of potential violations of the standard aforementioned assumptions for a case$^2$ study. For example, 1) according to the World Health Organization, violent behavior may increase the risk of unintentional injuries, potentially leading to accidental deaths; and 2) violent behavior can complicate the classification of deaths, making it challenging to distinguish between accidental deaths and suicides, as seen in cases involving driver suicides.

In a later study, \citet{small2013case} suggested that, under specific simple models, a more restrictive case definition information with fewer cases could result in increased design sensitivity, leading to reduced sensitivity in larger samples. The proposal of matching narrow cases to multiple referents might be equally cost-effective compared to pair matching with a broader case definition, while also potentially being less susceptible to unmeasured biases. 

In a subsequent work, \citet{ye2021combining} introduced a novel sensitivity analysis framework for integrating broad and narrow case definitions in matched case-control studies. This framework addresses both unmeasured confounding and selection bias simultaneously, enhancing analytical rigor. Their approach includes a robust randomization-based testing procedure using narrow case matched sets.

More recently, \citet{chen2022using} proposed a novel approach to mitigate sensitivity to hidden bias when conducting statistical inference on the attributable fraction \citep{walter1978calculation,poole2015history} by utilizing case definition information, such as cancer subtypes. This approach acknowledges that the treatment may affect different case types differently. The authors investigate how leveraging case description information can reduce sensitivity to bias from unmeasured confounding through design sensitivity. They also consider the potential introduction of additional selection bias through an extra sensitivity parameter when leveraging case definition information.

The rest of this article is organized as follows. Section \ref{sec: review} introduces the setting and notation as well as a review of the attributable effect and a sensitivity analysis framework for unmeasured confounding in case$^2$-studies. Section \ref{sec: method} proposes our sensitivity analysis framework for attributable effects. We implement our proposed method with our motivating example in Section \ref{sec: case-studies}. Section \ref{sec: conclusion} presents a discussion. Proofs can be found in the supplementary material. Code is available on \url{https://github.com/dosen4552/case_2_studies}.

\section{Notation and Review: Attributable Effect and Sensitivity Analysis for Hidden Bias in Matched Case$^2$ Studies} 
\label{sec: review}

\subsection{Notation and Setting of Matched Case$^2$-Studies, and Hypothesis of Interest}

We consider the setting of matched case$^2$ studies. In the sample of interest, there are $I$ matched sets where the narrow case in each matched set $i = 1, 2, \cdots, I$ is matched with $J - 1 \geq 1$ marginal case(s). We use $ij, i = 1, \cdots, I, j = 1, \cdots, J$ to index the $j$-th unit in the $i$-th matched set, and $I \times J = N$. Let $\mathbf{x}_{ij}$ and $ Z_{ij}$ denote the observed covariates and the treatment indicator for unit $ij$, respectively.

Unit $ij$ has two potential outcomes, $r_{T_{ij}}$ and $r_{C_{ij}}$. They are the outcomes that unit $ij$ would have under the treatment and control, respectively. Therefore, $r_{T_{ij}}$ is observed if $ Z_{ij} = 1$ and $r_{C_{ij}}$ is observed if $ Z_{ij} = 0$ (\citet{fisher1923studies}; Neyman (1923)). We further denote the observed outcome $R_{ij} = Z_{ij} r_{T_{ij}} + (1 - Z_{ij}) r_{C_{ij}}$, and let $Z_{i^+} = \sum_{j=1}^J Z_{ij}$ be the number of treated units in matched set $i$. If a unit has $R_{ij} = 1$, we call this a broad case. The narrow case definition $\kappa_n (\cdot)$ is a function that map from a potential outcome to $\{0,1\}$. The observed narrow case outcome is denoted by $\kappa_n(R_{ij}) = Z_{ij} \kappa_n(r_{T_{ij}}) + (1 - Z_{ij}) \kappa_n(r_{C_{ij}})$, and a narrow case must be a broad case (i.e., $\kappa_n(R_{ij}) = 1$ implies $R_{ij} = 1$). Therefore, $R_{ij} = 0$ denotes a referent, $\kappa_n(R_{ij}) = 1$ a narrow case, and $R_{ij} = 1$ and $\kappa_n (R_{ij}) = 0$ a marginal case.  In matched case$^2$-studies, every unit is a broad case and there is no referent. With this notation, individual $ij$ has potential case variables $(r_{T_{ij}}, \kappa_n(r_{T_{ij}}) )$ if exposed to treatment, and potential case variables $(r_{C_{ij}}, \kappa_n(r_{C_{ij}}) )$ if not exposed to treatment, so the observed case variables are $(R_{ij}, \kappa_n(R_{ij}) )$. From our matching scheme, $\kappa_n(R_{i1}) = 1$, $R_{i1} = R_{i2} = \cdots = R_{iJ} = 1$, and $\kappa_n (R_{i2}) = \cdots = \kappa_n(R_{iJ}) = 0$, for $i = 1, \cdots, I$.

\begin{figure}[t]
	\centering
 \resizebox{!}{.25\textwidth}{
	\begin{tikzpicture}
		\node[state] (1) {$Z$};
		\node[state] (2) [below =of 1,xshift=3cm] {$\kappa_n(R)$};
		\node[state] (5) [above =of 2, yshift=1cm] {$X, U$};
		\path (1) edge node[above] {} (2);
		\path (5) edge node[el,above] {} (2);
		\path (5) edge node[el,above] {} (1);
	\end{tikzpicture}} 
	\caption{A directed acyclic graph (DAG) under $ H_0$ defined in (\ref{equ:hypothesis}), where $ Z $ is the treatment and $ \kappa_n(R) $ is the observed narrow case status, $ X $ is the vector of measured covariates, $ U $ is the vector of unmeasured confounders. Testing $H_0$ (\ref{equ:hypothesis}) within always-cases may suffer from selection bias issue. Always-case is defined with respect to the broad case definition and refers to a subject who would be a broad case regardless of receiving the treatment or not.  } 
 \label{fig: dag}
\end{figure}
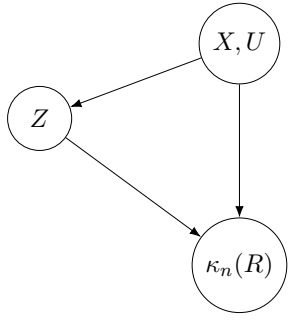

The vectors $\bm Z = \{Z_{11},\cdots,Z_{IJ} \}$ and $\bm R = (R_{11}, \cdots, R_{IJ})$ are both observed and they represent the treatment assignment and the observed outcome of all units. $\mathcal{Z}$ is the set of all possible values of $\bm Z$ with $\sum_{j=1}^J Z_{ij} = Z_{i^+}$ for $i = 1, \cdots, I$ and $|\mathcal{Z}| = \prod_{i=1}^I \binom{J}{Z_{i^+}}$. Also, we denote $u_{ij}$ as an unobserved normalized covariate for unit $ij$ (i.e. $0 \leq u_{ij} \leq 1$), $\bm u_i = (u_{i1}, u_{i2}, \cdots, u_{iJ})$  and $\mathcal{F} = \{(r_{T_{ij}},r_{C_{ij}},\mathbf{x}_{ij}, u_{ij}), i = 1, \cdots, I, j = 1, \cdots, J \}$ as the set of observed and unobserved covariates and outcomes. We assume matching has controlled the observed covariates in the sense that $\mathbf{x}_{ij} = \mathbf{x}_{ij'}$ for $1 \leq j < j' \leq J$ in each matched set $i$, but matched subjects could differ in their unobserved normalized covariates $u_{ij}$. The distribution of matched set $i$ could also be demonstrated in the top table of Table $\ref{tab: dist}$.  Figure \ref{fig: dag} demonstrates the basic idea of our settings. 

Throughout this paper, we follow the same assumption as \citet{rosenbaum2001effects,rosenbaum2002attributing}; that is, the treatment has nonnegative effect (i.e., $r_{T_{ij}} \geq r_{C_{ij}} $ and $ \kappa_n (r_{T_{ij}} ) \geq \kappa_n (r_{C_{ij}}) $). In our motivating example,ent is having violent behavior in the last year of life, so the nonnegative treatment effect assumption is saying that having violent behavior in the last year never prevents suicide death. Further discussion of the violation of nonnegative effect assumption can be found in the supplementary material.

Next, we consider the sign-score statistic which is defined below: 

\begin{definition}
A sign-score statistic for a binary outcome has the following form
\begin{align*}
        T =  \sum_{i=1}^I \sum_{j = 1}^{J} \kappa_n (R_{ij}) Z_{ij} = \sum_{i = 1}^I  C_i, \text{\quad where  $C_i = \sum_{j=1}^{J}\kappa_n (R_{ij})Z_{ij}$}.
\end{align*}
\end{definition}

 The sign-score statistic can be viewed as the number of treated narrow cases. This is the sum of the Subtable 1 of Table \ref{tab: dist} across all stratum $i$. The null hypothesis of interest in this article is 
\begin{equation}  \label{equ:hypothesis}
        H_0: r_{T_{ij}} - r_{C_{ij}} = 0 \quad \text{for all } i, j.
\end{equation} 
 This is the Fisher's sharp null hypothesis of no treatment effect based on the broad case definition.

In our motivating example, the sign-score statistic is the number of suicide deaths who had violent behavior in the last year of life, and the hypothesis of interest cares about the effect of having violent behavior in the last year of life on the risk of deaths including accidental deaths and suicide deaths.

\subsection{Review of the Attributable Effect in Case$^2$-Studies}

The attributable effect is a way to measure the magnitude of effect of a treatment on a binary outcome. It is denoted by $A$ and defined by $A  = \sum_{i=1}^I \sum_{j=1}^J Z_{ij}(\kappa_n (r_{T_{ij}}) - \kappa_n (r_{C_{ij}}) )$, that is the number of treated narrow cases actually caused by exposure to the treatment \citep{rosenbaum2001effects, rosenbaum2010design,rosenbaum2005attributable}. In our motivating example, the attributable effects are the number of suicide cases among people who had violent behavior in the last year of life that would not have occurred if those people did not have violent behavior in the last year of life.

\begin{table}[!htbp]

\centering
\begin{tabular}{|| c c c c ||}
\hline
 Case Type & Treated $Z_{ij} = 1$ & Control $Z_{ij} = 0$ & Total\\
 \hline
 \hline
$h_{ij} = 1$ & $\sum_{j} Z_{ij} h_{ij} R_{ij}$ & $ \sum_j(1 - Z_{ij}) h_{ij} R_{ij}$ & $\sum_j h_{ij} R_{ij}$ \\ 
 \hline
 $h_{ij} = 0$ & $\sum_j Z_{ij} (1 - h_{ij}) R_{ij}$ & $\sum_j (1 -Z_{ij}) (1 - h_{ij}) R_{ij}$ & $\sum_j (1 - h_{ij})R_{ij}$ \\
 \hline
 Total & $\sum_j Z_{ij} R_{ij}$ & $\sum_j (1 - Z_{ij})R_{ij}$ & $\sum_j R_{ij}$ \\
 \hline
 \hline
 Case Type & Treated $Z_{ij} = 1$ & Control $Z_{ij} = 0$ & Total\\
 \hline
 \hline
$h_{ij} = 1$ & $\sum_j Z_{ij} h_{ij} r_{C_{ij}}$ & $ \sum_j(1 - Z_{ij}) h_{ij} r_{C_{ij}}$ & $\sum_j h_{ij} r_{C_{ij}}$ \\ 
 \hline
 $h_{ij} = 0$ & $\sum_j Z_{ij} (1 - h_{ij}) r_{C_{ij}}$ & $\sum_j (1 -Z_{ij}) (1 - h_{ij}) r_{C_{ij}} $ & $\sum_j (1 - h_{ij}) r_{C_{ij}}$ \\
 \hline
 Total & $\sum_j Z_{ij} r_{C_{ij}}$ & $\sum_j (1 - Z_{ij}) r_{C_{ij}}$ & $\sum_j  r_{C_{ij}}$ \\
 \hline
 \hline
  Case Type & Treated $Z_{ij} = 1$ & Control $Z_{ij} = 0$ & Total\\
 \hline
 \hline
$h_{ij} = 1$ & $\sum_{j} Z_{ij} h_{ij} R_{ij} - A_0$ & $ \sum_j(1 - Z_{ij}) h_{ij} R_{ij}$ & $\sum_j h_{ij} R_{ij} - A_0$ \\ 
 \hline
 $h_{ij} = 0$ & $\sum_j Z_{ij} (1 - h_{ij}) R_{ij}$ & $\sum_j (1 -Z_{ij}) (1 - h_{ij}) R_{ij}$ & $\sum_j (1 - h_{ij})R_{ij}$ \\
 \hline
 Total & $\sum_j Z_{ij} R_{ij} - A_0$ & $\sum_j (1 - Z_{ij})R_{ij}$ & $\sum_j R_{ij}$ \\
 \hline
\end{tabular}
\caption{Top (Subtable 1): distribution of cases type by treatment for stratum $i$; Middle (Subtable 2): unobserved potential outcomes that would have been seen if the treated subjects had been spared the treatment for stratum $i$; Bottom (Subtable 3): observed distribution of cases type adjusted for the hypothesis for stratum $i$.}
\label{tab: dist}
\end{table}

Let $\bm \delta = (\delta_{11}, \cdots, \delta_{IJ})$ be the vector of treatment effects where $\delta_{ij} = r_{T_{ij}} - r_{C_{ij}}$. Consider a particular null hypothesis, $H_0: \bm \delta = \bm \delta_0$ for some specific vector $\bm \delta_0 = (\delta_{0_{11}}, \cdots, \delta_{0_{IJ}})$. We call $H_0: \bm \delta = \bm \delta_0$ ``incompatible'' if it is logically impossible given what is observed and assumed; otherwise, the hypothesis is ``compatible'' \citep{rosenbaum2001effects}. To understand ``compatible'', since $R_{ij} = Z_{ij} r_{T_{ij}} + (1 - Z_{ij}) r_{C_{ij}}$, if $Z_{ij} = 1$ and $R_{ij} = 0$, then $r_{T_{ij}} = r_{C_{ij}} = 0$ because $r_{T_{ij}} \geq r_{C_{ij}}$. Hence, $\delta_{0_{ij}} = 0$. Similar reason for $Z_{ij} = 0$ and $R_{ij} = 1$ because $r_{T_{ij}} = r_{C_{ij}} = 1$.  If the sharp null hypothesis $\bm \delta = \bm \delta_0$ is incompatible, then reject it (the type I error rate for incompatible hypotheses is then zero).  If the null hypothesis  $H_0: \bm \delta = \bm \delta_0$ is true, then $r_{C_{ij}}$ will be known for all $i, j$ since $r_{C_{ij}} = R_{ij} - Z_{ij} \delta_{0_{ij}}$. To fix idea, let $A_0 = \sum_{i=1}^I\sum_{j=1}^J Z_{ij} h_{ij} \delta_{0_{ij}}$ where  $h_{ij} = 1$ and $h_{ij} = 0$ represent narrow case and marginal case for unit $ij$, respectively, then under the null hypothesis, $T - A_0 = \sum_{i=1}^I \sum_{j=1}^J Z_{ij} h_{ij}r_{C_{ij}}$. If the hypothesis $H_0: \bm \delta = \bm \delta_0$ is true, then the Subtable 3 of Table \ref{tab: dist} equals the Subtable 2 of Table \ref{tab: dist}, which has the $\chi^2$ distribution with degree of freedom 1, so the hypothesis can be tested by applying McNemar-Mantel-Haenszel test to the Subtable 3 of Table \ref{tab: dist}, that is, by comparing this table to the $\chi^2$ distribution with degree of freedom 1 distribution. A compatible hypothesis $H_0: \bm \delta = \bm \delta_0$ is tested by assuming the hypothesis for the purpose of testing it, and applying the McNemar-Mantel-Haenszel test $T = \sum_{i=1}^I \sum_{j=1}^J Z_{ij} h_{ij} r_{C_{ij}}$ to the adjusted outcomes $ h_{ij} r_{C_{ij}} = h_{ij} - Z_{ij} h_{ij} \delta_{0_{ij}}$. Under $H_0$, if the probability of a unit getting treated is constant in matched sets, $\theta_{ij} = \theta_{ij'}$ for each $i, j, j'$ where $\theta_{ij} = \mathbb{P} (Z_{ij} = 1 | \mathcal{F}, \mathcal{Z})$, then the conditional distribution of $Z_{ij}$ given $Z_{i^+}$ does not depend on $\theta_{ij}$, and $T$ is that of sum of $I$ independent Bernoulli variables with probabilities of success $\mathbb{P}(r_{C_{i1}} Z_{i1} = 1 | Z_{i^+}) =  r_{C_{i1}} Z_{i^+}/J$ since $h_{i1} = 1$ and $h_{ij} = 0$ for $j = 2, \cdots, J $ in our settings \citep{cox1966simple}. If a compatible hypothesis, $H_0: \bm \delta = \bm \delta_0$, attributes the first case in matched set $i$ to an effect of the treatment, that is, if $Z_{ij} h_{ij} \delta_{0_{ij}} = 1$ so that $h_{ij} R_{ij} = 1$ but $h_{ij} r_{C_{ij}} = h_{ij} - Z_{ij} h_{ij} \delta_{0_{ij}} = 0$, then matched set $j$ becomes concordant because $\pi_{i1} = r_{C_{i1}} Z_{i1}/J = 0$, which is, for all practical purposes, the same as deleting matched set $j$. 

Note that the attributable effect, $\sum_{i,j} Z_{ij} h_{ij} \delta_{ij}$, is a random variable rather than a fixed parameter, precluding direct hypothesis testing about its value. Nevertheless, statistical inference on the attributable effect can still be conducted by constructing a one-sided $(1 - \alpha) \times 100\%$ prediction interval. This interval has the interpretation that, in repeated studies, the attributable effect will fall within the interval $(1-\alpha) \times 100\%$ of the time.

To construct one-sided prediction interval for $A$, we start with $A = A_0 = 0$, then conduct McNemar-Mantel-Haenszel test on the  Subtable 3 of Table \ref{tab: dist}. If we reject the null, increasing $A_0$ by 1 until fail to reject the null. Say if we obtain $A_0 = A_*$, then the one-sided prediction interval for $A$ is $\{A: A \geq A_* \}$.

%\begin{table}[!htbp]

%\centering
%\begin{tabular}{|| c c c c ||}
%\hline
% Case Type & Treated $Z_{ij} = 1$ & Control $Z_{ij} = 0$ & Total\\
% \hline
% \hline
%$h_{ij} = 1$ & $\sum_{j} Z_{ij} h_{ij} R_{ij} - A_0$ & $ \sum_j(1 - Z_{ij}) h_{ij} R_{ij}$ & $\sum_j h_{ij} R_{ij} - A_0$ \\ 
% \hline
% $h_{ij} = 0$ & $\sum_j Z_{ij} (1 - h_{ij}) R_{ij}$ & $\sum_j (1 -Z_{ij}) (1 - h_{ij}) R_{ij}$ & $\sum_j (1 - h_{ij})R_{ij}$ \\
% \hline
% Total & $\sum_j Z_{ij} R_{ij} - A_0$ & $\sum_j (1 - Z_{ij})R_{ij}$ & $\sum_j R_{ij}$ \\
% \hline
%\end{tabular}
%\caption{Observed distribution of cases type adjusted for the hypothesis for stratum $i$}
%\label{tab: adjust}
%\end{table}

%If we further denote $B_i = \sum_{j=1}^J  Z_{ij} h_{ij} r_{C_{ij}}$ and $r_{C_{i^+}} = \sum_{j=1}^J r_{C_{ij}}$ for the number who would have had events if exposure to treatment had been prevented, then

%\begin{align*}
  %\tilde{\pi}_i = \frac{ Z_{i^+} r_{C_{i^+}}  }{ Z_{i^+} r_{C_{i^+}} +  \Gamma (J - Z_{i^{+}} r_{C_{i^+}} )  }  \leq \mathbb{P} (B_i = 1) \leq \frac{\Gamma Z_{i^+} r_{C_{i^+}}  }{ \Gamma Z_{i^+} r_{C_{i^+}} + J - Z_{i^{+}} r_{C_{i^+}}  }  = \tilde{\tilde{\pi}}_i,
%\end{align*}

%and, in particular, in a randomized experiment when $\Gamma = 1$, $\mathbb{P} (B_i = 1) =Z_{i^+} r_{C_{i^+}}/J$. 

 \subsection{Sensitivity Analysis: Hidden Bias} \label{subsec: sens}

 In observational studies, if a study is free of hidden bias, then two units with the same observed covariates have the same probability of receiving treatment. There is \emph{hidden bias} if two units with the same covariates have different probability of receiving treatment (i.e. $\mathbf{x}_{ij} = \mathbf{x}_{ij'}$ but $\mathbb{P} (Z_{ij} = 1 | \mathcal{F}, \mathcal{Z}) \neq \mathbb{P}(Z_{ij'} = 1 | \mathcal{F},\mathcal{Z})$ for $i = 1, \cdots, I, 1 \leq j < j' \leq J$). A sensitivity analysis asks how the causal conclusion would be altered by hidden bias of various magnitudes. We adapt the sensitivity analysis framework from \citet{rosenbaum2002overt}.  The odds that unit $ij$ and unit $ij'$ receive the treatment are $\theta_{ij} / (1 - \theta_{ij})$ and $\theta_{ij'} / (1 - \theta_{ij'})$, respectively. We consider a sensitivity parameter $\Gamma \geq 1$ that bounds the amount of hidden bias:
\begin{equation} \label{odds: hidden bias}
    \frac{1}{\Gamma} \leq \frac{\theta_{ij} (1 - \theta_{ij'})}{\theta_{ij'} (1 - \theta_{ij})} \leq \Gamma \qquad \text{for $i = 1, \cdots, I, 1 \leq j < j' \leq J$ with $\mathbf{x}_{ij} = \mathbf{x}_{ij'}$}. 
\end{equation}
Here, $\Gamma$ is a measure of the degree of departure from a study that is free of hidden bias. \citet{rosenbaum2002overt} also proposed the sensitivity model
\begin{equation} \label{model: hidden bias}
    \log (\frac{\theta_{ij}}{1 - \theta_{ij}}) = \alpha (\mathbf{x}_{ij}) + \log (\Gamma) u_{ij} \qquad \text{with $0 \leq u_i \leq 1$}
\end{equation}
where $\alpha (\cdot)$ is an unknown function. Proposition 12 in Chapter 4 from \citet{rosenbaum2002overt} states that (\ref{odds: hidden bias}) is equivalent to the existence of model (\ref{model: hidden bias}). Note that when $\Gamma = 1$, unit $ij$ and unit $ij'$ with the same observed covariates have the same odds of receiving treatment, so there is no hidden bias. Recall that under the null hypothesis $H_0: \bm \delta = \bm \delta_0 $, $r_{C_{ij}} = R_{ij} - Z_{ij} \delta_{0_{ij}}$ and $T - A_0 = \sum_{i=1}^I \sum_{j=1}^J Z_{ij} \kappa_n (r_{C_{ij}})$. If we denote $B_i = \sum_{j} Z_{ij} \kappa_n (r_{C_{ij}}) $ which is the number of treated narrow cases not caused by exposure to the treatment in matched set $i$, and $\kappa_n(r_{C_{i^+} }) = \sum_{j}\kappa_n (r_{C_{ij}})$ which is the number of untreated narrow cases in matched set $i$,  then we will have:

\begin{equation} \label{equ: rosenbaum sens}
       \frac{ Z_{i^+} \kappa_n(r_{C_{i^+} }) }{ Z_{i^+} \kappa_n(r_{C_{i^+} }) + \Gamma \left(J - Z_{i^+} \kappa_n(r_{C_{i^+} }) \right) } \leq \mathbb{P} (B_i = 1) \leq  \frac{ \Gamma Z_{i^+} \kappa_n(r_{C_{i^+} }) }{ \Gamma Z_{i^+} \kappa_n(r_{C_{i^+} }) +  J - Z_{i^+} \kappa_n(r_{C_{i^+} }) } = \bar \pi_{i},
\end{equation}
and in particular, in a randomized experiment, $\Gamma = 1$, so that $ \mathbb{P} (B_i = 1) =  Z_{i^+} \kappa_n(r_{C_{i^+} } ) / J$. Under the null hypothesis of (\ref{equ:hypothesis}), in a sensitivity analysis, the upper bound of the tail area $\mathbb{P} (\sum_{i} B_i \geq k ) $ can be approximated using normal approximation \citep{rosenbaum2002attributing}:
\begin{align*}
  1 - \Phi ( \frac{k - \sum_i \bar \pi_i}{\sqrt{ \sum_i \bar \pi_i (1 - \bar \pi_i)  }  }  ). 
\end{align*}

%\textcolor{blue}{\citet{gastwirth2000asymptotic} demonstrates one can find the $\bm \delta_0$ with $a = \sum_{i,j}Z_{ij}h_{ij}\delta_{0_{ij}}$ that maximize the expectation $\sum_{i} \bar \pi_{i}$, and if several $\bm \delta_0$ do this, then finds among these the one that also maximizes the variance $\sum_i \bar \pi_{i} (1- \bar \pi_{i})$. And this exhibits asymptotic separability which} implies that in the presence of numerous matched sets, individual optimizations can be conducted within each matched set, and subsequently, the results can be aggregated with negligible error. As a result, this would be upper bound of the one-sided significant level for testing the null hypothesis (\ref{equ:hypothesis}) when it is computed at the observed value of the statistic, $k = T - A_0$. Detail discussion can be found in Section 3.1 in \citet{gastwirth2000asymptotic} and Section 3.4 in \citet{rosenbaum2002attributing}. 

\citet{gastwirth2000asymptotic} demonstrates that one can find the asymptotically correct upper bound of the one-sided significance level by doing individual optimizations within each matched set. Specifically, one can find the $\bm \delta_0$ with $a = \sum_{i,j}Z_{ij}h_{ij}\delta_{0_{ij}}$ that maximize the expectation $\sum_{i} \bar \pi_{i}$, and the variance $\sum_i \bar \pi_{i} (1- \bar \pi_{i})$. \citet{gastwirth2000asymptotic} showed that under regularity conditions, as the number of matched sets goes to infinity, the difference between the upper bound of the one-sided significance level for this  $\bm \delta_0$ and the true upper bound converges to 0. Detailed discussion can be found in Section 3.1 in \citet{gastwirth2000asymptotic} and Section 3.4 in \citet{rosenbaum2002attributing}.

\section{A Sensitivity Analysis Framework in Case$^2$-Studies}
\label{sec: method}

\subsection{Sensitivity Analysis Model in Case$^2$-Studies}

Assuming that $Z_{ij} \indep (r_{T_{ij}}, r_{C_{ij}}, \kappa_n(r_{T_{ij}}), 
\kappa_n(r_{C_{ij}}) ) | x_{ij}, u_{ij}$ for all $i,j$, and independence among distinct subjects, we propose a testing procedure for the null hypothesis in case$^2$ studies, inspired by the sensitivity model for case-control studies by \citet{ye2021combining}. This procedure accounts for violation of the assumptions 1.) treatment does not cause marginal case, 2.) treatment does not alter individual's case type, and 3.) there is no unmeasured confounders by considering:

\begin{equation}
\label{sensitivity model}
    \begin{split}
        \pi_{ij} &:= \mathbb{P}(Z_{ij} = 1|\mathbf{x}_{ij}, u_{ij}) = \text{expit}(\alpha_z(x_{ij}) + \log(\Gamma) u_{ij} ) \\
        \theta_{T_{ij}} &:= \mathbb{P} (\kappa_n(r_{T_{ij}}) = 1 | \mathbf{x}_{ij}, u_{ij}, r_{T_{ij}} = r_{C_{ij}}  = 1  ) \\
        \theta_{C_{ij}} &:= \mathbb{P} (\kappa_n(r_{C_{ij}}) = 1 | \mathbf{x}_{ij}, u_{ij}, r_{T_{ij}} = r_{C_{ij}}  = 1  ) \\
        \text{where } & 0 \leq u_{ij} \leq 1, 1 \leq  \frac{\theta_{T_{ij}}}{\theta_{C_{ij}}}  \leq \Theta,  1   \leq \frac{1 - \theta_{C_{ij}}}{1 - \theta_{T_{ij}}} \leq \Delta,\\
        &0 < \theta_{T_{ij}} < 1, 0 < \theta_{C_{ij}} < 1, \text{and } \Theta, \Gamma, \Delta \geq 1.  
    \end{split}
\end{equation}

In the sensitivity model \eqref{sensitivity model}, we introduce $\Gamma, \Theta$ and $\Delta$ as the sensitivity parameters to assess the impact of unmeasured confounders, selection bias, and the deviation of the assumption that treatment would not cause marginal cases, respectively. Specifically, $\Gamma=1$ signifies the absence of unmeasured confounders.  The parameter $\Theta, \Delta$ set bounds on the risk ratios, $\theta_{T_{ij}}/\theta_{C_{ij}}$ and $(1 - \theta_{C_{ij}})/(1 - \theta_{T_{ij}})$, reflecting the likelihood of a case being categorized as narrow or marginal, with or without treatment, in case$^2$-studies. Our model extends \citet{ye2021combining}'s by imposing an additional constraint, $1 \leq (1 - \theta_{C_{ij}})/(1 - \theta_{T_{ij}}) \leq \Delta$, tailored to case$^2$-studies. This constraint indicates that having violent behavior in the last year of life could make a case at least $1/\Delta$ times more likely to an accidental death case. Under Fisher's null hypothesis \eqref{equ:hypothesis}, $\Theta=1$ indicates that treatment does not shift a case's categorization, thus no selection bias. Conversely, deviations from $\Theta=1$ quantify the degree of selection bias. $\Delta = 1$ indicates that treatment does not cause marginal cases. When $\Theta = 1$ and $\Delta = 1$, then under Fisher's null $H_0$ in (\ref{equ:hypothesis}), the treatment has no effect on the narrow case definition (i.e., $\kappa_n(R_{ij}) \indep Z_{ij} \mid \mathcal{F}$ under $H_0$). This triple-parameter sensitivity model uniquely allows for the simultaneous consideration of unmeasured confounding, selection bias, and the deviation of the assumption that treatment would not cause marginal cases with a provision for adjusting the model by setting $ \Theta^{-1} \leq \theta_{Tij}/\theta_{Cij} \leq \Theta$ for $\Theta \geq 1$ and $ \Delta^{-1} \leq (1 - \theta_{C_{ij}})/(1 - \theta_{T_{ij}}) \leq \Delta$ for $\Delta \geq 1$. Moreover,  $u_{ij}$ is a unmeasured confounder that satisfies the condition $ Z_{ij} \indep ( r_{T_{ij}}, r_{C_{ij}}, \kappa_n(r_{T_{ij} }), \kappa_n(r_{C_{ij} })) | x_{ij}, u_{ij} $. But this  $u_{ij}$  needs not modify  $\theta_{T_{ij}}/\theta_{C_{ij}}$ and $ (1 - \theta_{C_{ij}})/(1 - \theta_{T_{ij}})$. Additionally,
$\theta_{T_{ij}}/\theta_{C_{ij}}$ and $ (1 - \theta_{C_{ij}})/(1 - \theta_{T_{ij}})$ can be rewritten as 

\begin{align*}
    	\frac{\theta_{T_{ij}}}{\theta_{C_{ij}}}&= \frac{\mathbb{P}(\kappa_n(R_{ij})=1\mid  x_{ij}, u_{ij}, r_{C_{ij}}=  r_{T_{ij}}= 1,  Z_{ij} = 1)}{\mathbb{P}(\kappa_n(R_{ij})=1\mid  x_{ij}, u_{ij}, r_{C_{ij}}=  r_{T_{ij}}= 1, Z_{ij}=0 )}  \nonumber\\
	&  = \frac{\text{Odds} \{ \mathbb{P}(Z_{ij}= 1\mid x_{ij}, u_{ij}, r_{C_{ij}}= r_{T_{ij}}= 1, \kappa_n(R_{ij}) = 1)\}  }{\text{Odds} \{ \mathbb{P}(Z_{ij}= 1\mid x_{ij}, u_{ij}, r_{C_{ij}}=r_{T_{ij}}= 1)\}}\leq \Theta,  \\
   \frac{1 - \theta_{C_{ij}}}{1 - \theta_{T_{ij}}} &= \frac{\mathbb{P}(\kappa_n(R_{ij})=0\mid  x_{ij}, u_{ij}, r_{C_{ij}}= r_{T_{ij}}= 1,  Z_{ij} = 0)}{\mathbb{P}(\kappa_n(R_{ij})=0\mid  x_{ij}, u_{ij}, r_{C_{ij}}=  r_{T_{ij}}= 1, Z_{ij}=1 )} \\
   & = \frac{\text{Odds} \{ \mathbb{P}(Z_{ij}= 0\mid x_{ij}, u_{ij}, r_{C_{ij}}= r_{T_{ij}}= 1, \kappa_n(R_{ij}) = 0)\}  }{\text{Odds} \{ \mathbb{P}(Z_{ij}= 0\mid x_{ij}, u_{ij}, r_{C_{ij}}=r_{T_{ij}}= 1)\}} \leq \Delta, 
\end{align*}
where $ \text{Odds} \{p\} = p/(1-p)$. Thus, $ \theta_{Tij}/\theta_{Cij} $ is the odds ratio of  receiving treatment for the same subject with and without 
the information that this subject is a narrow case, which is different from the odds ratio of receiving treatment for two matched subjects in Rosenbaum's sensitivity model \eqref{equ: rosenbaum sens}. And $(1 - \theta_{C_{ij}})/(1 - \theta_{T_{ij}})$ is the odds ratio of not receiving treatment for the same subject with and without the information that this subject is a narrow case. 

\begin{figure}[!htbp]
	\centering
	\resizebox{!}{!}{
		\begin{tikzpicture}
			\draw[draw=black] (10,5) rectangle ++(6,1.5);
			\draw[draw=black] (17,5) rectangle ++(6,1.5);
			\draw[draw=black] (10,3) rectangle ++(6,1.5);
			\draw[draw=black] (17,3) rectangle ++(6,1.5);
			\node[above] at (13,6.5) {$Z_{ij} = 1$};
			\node[above] at (20,6.5) {$Z_{ij} = 0$};
			\node[right] at (8,5.8) {$h_{ij} = 1$};
			\node[right] at (8,3.8) {$h_{ij} = 0$};
			\node[right] at (10.2,3.7) {$ \{ij\mid r_{T_{ij}}= 1, \kappa_n(r_{T_{ij}})= 0 \} $ };
			\node[right] at (17.2,3.8) {$ \{ij\mid r_{C_{ij}}= 1, \kappa_n(r_{C_{ij}})= 0 \} $ };
			\node[right] at (11.4,5.8) {$ \{ij\mid \kappa_n(r_{T_{ij}})= 1 \} $ };
			\node[right] at (18.4,5.8) {$ \{ij\mid \kappa_n(r_{C_{ij}})= 1 \} $ };
			\draw[->] (18,5.3) -- (15,4.2) node[pos=1.1,right] {$ h_{NM} $};
			\draw[->] (18,4.2) -- (15,5.3) node[pos=1.2,right] {$ h_{MN} $};
			\draw[->] (18,6.2) -- (15,6.2) node[pos=1.3,right] {$ h_{NN} $};
			\draw[->] (18,3.3) -- (15,3.3) node[pos=1.33,right] {$ h_{MM} $};
	\end{tikzpicture}}
	\caption{An illustration of how the cases are moved by the treatment in case$^2$-studies, within the stratum defined by levels of $( x_{ij}, u_{ij}) $ in which  
		$ \theta_{Tij}/\theta_{Cij} $ and $(1 - \theta_{C_{ij}})/(1 - \theta_{T_{ij}})$ are the largest. The left panel is  the case allocation when all are treated, and the right panel is the case allocation when all the subjects are untreated.  The numbers $ h_{NN}, h_{MN}, h_{NM}, h_{MM} $ are expected values. }
  \label{fig: move}
\end{figure}
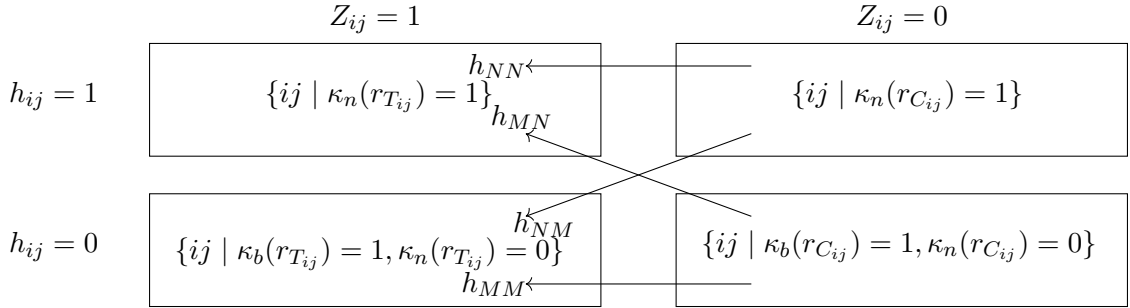

To interpret $\Theta$ and $\Delta$ within the stratum defined by the levels of \( (x_{ij}, u_{ij}) \) where $\theta_{T_{ij}}/\theta_{C_{ij}}$ and $(1 - \theta_{C_{ij}})/(1 - \theta_{T_{ij}})$ attain their maximum, we use Figure \ref{fig: move} to demonstrate the impact of change of treatment status on the cases. The left panel illustrates case allocation with treated units, while the right panel depicts allocation with untreated units. Figure \ref{fig: move} reveals that $h_{NN}$ narrow cases and $h_{MM}$ marginal cases remain unaffected by the treatment. Conversely, the treatment induces a shift, transforming \( h_{MN} \) marginal cases into narrow cases and \( h_{NM} \) narrow cases into marginal cases, where \( h_{NN} \), \( h_{MN} \), \( h_{NM} \), and \( h_{MM} \) denote expected values. Consequently, $\theta_{T_{ij}}/\theta_{C_{ij}} \leq \Theta$ and $(1 - \theta_{C_{ij}})/(1 - \theta_{T_{ij}}) \leq \Delta$ need $(h_{NN}+h_{MN}) /(h_{NN}+h_{NM})\leq \Theta$ and $ (h_{MN}+h_{MM})/(h_{NM}+h_{MM}) \leq \Delta$, which can be expressed as  $(h_{MN} - h_{NM})/(h_{NN} + h_{MN}) \leq 1 - 1/\Theta$ and $(h_{MN} - h_{NM})/(h_{MN} + h_{MM}) \leq 1 - 1/\Delta$. In the settings of case$^2$-studies, When it is reasonable to assume that $h_{NM}$ is small, $(h_{MN} - h_{NM})/(h_{NN} + h_{MN})$ and $(h_{MN} - h_{NM})/(h_{MN} + h_{MM})$ can be further approximated by $h_{MN}/(h_{NN} + h_{MN}) \leq 1 - 1/\Theta $ and $h_{MN}/(h_{MN} + h_{MM}) \leq 1 - 1/\Delta$. In the context of our application, $\Theta = 1.10$ means that up to $9\%$ of the units completingsuicide when having violent behavior in the last year of life would encounter accidental death if not having violent behavior, and $\Delta = 1.10$ means that up to $9\%$ of the accidental deaths due to having violent behavior in the last year of life. For a comprehensive exploration of sensitivity analysis calibration, we refer to works such as \cite{Imbens2003}, \cite{hsu2013calibrating}, \cite{cinelli2020making}, and \cite{zhang2020calibrated}.

\subsection{Inference for Attributable Effects}

Denote $\mathcal{B}$ is the event that the first element in stratum $i$ is the narrow case while the rest of them are marginal cases for $i = 1, \cdots, I$. To develop a valid hypothesis testing procedure for Fisher's sharp null \eqref{equ:hypothesis}, we first study $\mathbb{P} (C_i = 1 |\mathcal{F}, \mathcal{B}, \mathcal{Z} ) $. The following proposition places bounds on this probability.

\begin{proposition} \label{prop: bound}
    Under the null hypothesis \eqref{equ:hypothesis} and the sensitivity model \eqref{sensitivity model} with fixed $\Gamma, \Theta$ and $\Delta$, then
\begin{equation} \label{equ: bound Bi}
     \frac{Z_{i^+}}{Z_{i^+} + (J - Z_{i^+}) \Gamma } \leq \mathbb{P} (C_i = 1 | \mathcal{F}, \mathcal{B}, \mathcal{Z}) \leq \frac{Z_{i^+} \Theta \Delta \Gamma }{Z_{i^+} \Theta \Delta \Gamma + (J - Z_{i^+})}.
\end{equation}
The lower bound is attained at $\bm u_i = (0,1,\cdots,1) $ and $(1 - \theta_{C_{ij}})/(1 - \theta_{T_{ij}})  = \theta_{T_{ij}}/ \theta_{C_{ij}} = 1$, and the upper bound is attained at $\bm u_i = (1,0,\cdots,0) $, $ \theta_{T_{ij}}/ \theta_{C_{ij}} = \Theta,$ and $(1 - \theta_{C_{ij}})/(1 - \theta_{T_{ij}}) = \Delta$. 
\end{proposition}

The proof of Proposition \ref{prop: bound} can be found in the Supplementary material. Proposition \ref{prop: bound} states that in the presence of hidden bias, selection bias, as well as the deviation from the assumption that the treatment does not cause marginal cases, the bounds of $\mathbb{P}(C_i = 1 | \mathcal{F}, \mathcal{B}, \mathcal{Z})$ can be expressed using sensitivity parameters $\Gamma$, $\Theta$, and $\Delta$. If the null hypothesis \eqref{equ:hypothesis} is true, then $\kappa_n(r_{C_{ij}})$ would be known for all $i,j$, because $\kappa_n(r_{C_{ij}})$ would equals the observed quantity, $\kappa_n(R_{ij}) = Z_{ij} h_{ij} \delta_{0_{ij}} $. Write $A_0 = \sum_{i,j} Z_{ij}h_{ij}\delta_{0_{ij}}$, then under the null hypothesis \eqref{equ:hypothesis}, $T - A_0 = \sum_{i,j} Z_{ij} h_{ij} r_{C_{ij}}$, and $\sum_{i,j} Z_{ij} h_{ij} r_{C_{ij}}$ is the sum of $I$ independent binary random variables. Recall that we denote $B_i = \sum_{j} Z_{ij} \kappa_n (r_{C_{ij}}) $ and $\kappa_n(r_{C_{i^+}}) = \sum_j \kappa_n (r_{C_{ij}})$. Consequently, motivated from Proposition \ref{prop: bound}, we have the following bounds for $\mathbb{P}(B_i = 1|\mathcal{F},\mathcal{B},\mathcal{Z})$ in matched case$^2$-studies because $\sum_{j}h_{ij} R_{ij} = 1$ for every $i = 1,2,\cdots,I$: 

\begin{align*}
      \frac{Z_{i^+} \kappa_n(r_{C_{i^+} }) }{Z_{i^+} \kappa_n(r_{C_{i^+} }) + \left(J - Z_{i^+} \kappa_n(r_{C_{i^+} })\right) \Gamma } \leq \mathbb{P} (B_i = 1 | \mathcal{F}, \mathcal{B}, \mathcal{Z}) \leq \frac{Z_{i^+} \kappa_n(r_{C_{i^+} }) \Theta \Delta \Gamma }{Z_{i^+}\kappa_n(r_{C_{i^+} }) \Theta \Delta \Gamma + \left( J - Z_{i^+} \kappa_n(r_{C_{i^+} })\right) } =\bar{\bar{\pi}}_{i}. 
\end{align*}

It is important to note that if the null is true, then $ \bar{\bar{\pi}}_{i} $ can be calculated from the observed data and the hypothesis. Next, we demonstrate a general procedure to find the attributable effect. The essential idea is to implement large-sample approximation to the exact inference via asymptotic separability \citep{gastwirth2000asymptotic}. The following steps find the $\bm \delta_0$ with $a =  \sum_{i,j} Z_{ij} h_{ij} \delta_{0_{ij}}$:

\begin{itemize}
 \item[1.] If $a \geq \sum_{i,j} Z_{ij} h_{ij} R_{ij}$, then $a$ or fewer of the treated subjects had events, whether caused by treatment or not, so it is certain that $a$ or fewer had events caused by the treatment; stop. Otherwise, if $\sum_{i,j}Z_{ij} h_{ij} R_{ij} > a$, then continue 2.
    \item[2.] Denote $L_{i^+} = \sum_{j = 1}^J h_{ij} R_{ij}$. For each matched set $i$ with $\sum_{j=1}^J Z_{ij} h_{ij} R_{ij} = 1$, calculate
    \begin{align*}
        \bar{\bar{\lambda}}_i &= \frac{\Theta^2 \Gamma  Z_{i^+} L_{i^+}}{\Theta^2 \Gamma  Z_{i^+}L_{i^+} + J - Z_{i^+}L_{i^+}} \\
        \bar{\lambda}_i &= \frac{\Theta^2 \Gamma Z_{i^+}(L_{i^+}-1)}{\Theta^2 \Gamma Z_{i^+}(L_{i^+}-1) + J - Z_{i^+}(L_{i^+}-1)} \\
        \bar{\bar{w}}_i &= \bar{\bar{\lambda}}_i (1 - \bar{\bar{\lambda}}_i) \\
        \bar{w}_i & =  \bar{\lambda}_i (1 -  \bar{\lambda}_i)
    \end{align*}
  \item[3.] Select exactly $a$ of the matched sets $i$ with $\sum_{j=1}^J Z_{ij} h_{ij} R_{ij} = 1$ having the smallest values of $\bar{\bar{\lambda}}_i - \bar{\lambda}_i$. If ties among the $\bar{\bar{\lambda}}_i - \bar{\lambda}_i$ mean that several different groups of $a$ matched sets all have the smallest values of $\bar{\bar{\lambda}}_i - \bar{\lambda}_i$, then among these several groups with the smallest $\bar{\bar{\lambda}}_i - \bar{\lambda}_i$, pick any one group with the smallest values of $\bar{\bar{\lambda}}_i - \bar{\lambda}_i$. For the selected $a$ matched sets, let $\bar{\bar{\pi}}_i = \bar{\lambda}_i$, whereas for the remaining $I - a$ matched sets, let $\bar{\bar{\pi}}_i =  \bar{\bar{\lambda}}_i$. If $a + \sum_{i} \bar{\bar{\pi}}_i \geq \sum_{i,j}Z_{ij} h_{ij} R_{ij}$, then there is one $\bm u$ and one $\bm \delta_0$ with $a = \sum_{i,j} Z_{ij} h_{ij} \delta_{0_{ij}}$ events caused by treatment that would lead us to expect more than the observed number $\sum_{i,j}Z_{ij} h_{ij} R_{ij}$ of events among treated subjects; conclude that this $a$ is plausible and stop; otherwise, continue with step 4.
  \item[4.] Calculate the large-sample approximation to the upper bound on the significance level 
  \begin{align*}
      1 - \Phi (\frac{ (T - a) - \sum_i \bar{\bar{\pi}}_i  }{\sqrt{\sum_{i} \bar{\bar{\pi}}_i (1 - \bar{\bar{\pi}}_i ) }}).
  \end{align*}
  If it is small (e.g., $< 0.05$), then conclude that every compatible $\bm \delta_0$ with $a \geq \sum_{i,j} Z_{ij} h_{ij} \delta_{0_{ij}}$ that it is not plausible that exposure to treatment caused $a$ or fewer events. 
\end{itemize}

\section{Case Study: Violence, Suicide Deaths, and Accidental Deaths}

\label{sec: case-studies}

\subsection{Analysis and Results}

In this case study, we aim to examine the connection between violent behavior in the last year of life and the associated suicide risk in comparison to accidental death after controlling for alcohol use disorders \citep{conner2001violence}. Specifically, we define subjects who are ``having violent behavior in the last year of life'' if subjects answered ``often'' or ``sometimes'' to this question: ``How often did --[the decedent subject]  make violent threats or attempts''. In addition, history of alcohol misuse was coded as positive if the informant endorsed one or more items on the CAGE questionnaire, a 4-item screening tool for detecting alcohol use disorders; it was coded as negative if no CAGE items were endorsed. In this study, we employ accidental deaths as the reference group, as it is possible that having violent behavior in the last year of life constitutes a risk factor for accidental death, hence the assumption that treatment does not cause marginal cases is potentially violated. Moreover, it is essential to acknowledge the potential presence of selection bias, as the occurrence of violent behavior could conceivably influence the categorization of deaths from accidental to suicidal. For instance, distinguishing between accidental death and suicide can be challenging, as seen in cases of driver suicides \citep{ohberg1997driver}. In the NMFS, respondents reporting on a deceased individual may mention threats of suicide or self-directed violence when queried about violent threats or attempts \citep{conner2001violence}. Consequently, such reports of violent behavior could shift the classification of a death from what might initially be deemed an accidental death (potentially a misclassified suicide) to a recognized suicide. Therefore, it is crucial to perform sensitivity analysis to account for hidden bias, selection bias, and the deviation from the assumption that treatment does not cause marginal cases in this case study.

\begin{table}[!htbp]
\centering
\caption{\small Demographic distribution of data example stratified by suicide death and accidental death from NMFS 1993.}
\label{tab:demographics}
\resizebox{0.8\textwidth}{!}{
\begin{tabular}{lcccc}
  \hline \multirow{3}{*}{\begin{tabular}{c}\textbf{ }\end{tabular}}
   & \multirow{4}{*}{\begin{tabular}{c} Suicide death \% \\  ($I = 1365$)\end{tabular}} &
   \multirow{4}{*}{\begin{tabular}{c} Accidental death \% \\ ($2I = 2730$)\end{tabular}} & \multirow{4}{*}{\begin{tabular}{c} Standardized \\  mean difference \end{tabular}}          &
 \\ \\ \\
  \hline
    %\textbf{Firearms in home} & 898 (65.8) & 1111 (40.7) & 	\\
    \textbf{Having violent behavior} & 247 (18.1) & 183 (6.7) & 	\\
    \textbf{Alcohol misuse} & 519 (38.0) & 774 (28.4) & 0.120	\\
    \textbf{Sex } \\
  \hspace{0.3cm} Male & 974(71.4) & 1948(71.4)  & 0.130\\ 
  \hspace{0.3cm} Female & 391(28.6)	 & 782(28.6)    & -0.130\\ 
    \textbf{Race } \\
  \hspace{0.3cm} White & 1209(88.6) & 2418(88.6) & 0.170\\ 
  \hspace{0.3cm} Black & 117(8.6) & 270(9.9)  & -0.190\\ 
  \hspace{0.3cm} Other & 39(2.86) & 42(1.5) & 0.010\\ 
   \textbf{Age at death, median (IQR)} & 46.0 (31, 67) & 41.0 (29, 65) & 0.030	\\
      \textbf{Lived alone} & 362(26.5) & 562(20.6) & 0.100	\\
        \textbf{Martial status} \\
  \hspace{0.3cm} Married & 588(43.1) & 1218(44.6) & -0.070 \\ 
  \hspace{0.3cm} Widowed & 178(13.0)	 & 315(11.5)    & 0.010\\ 
  \hspace{0.3cm} Divorced/separated & 209(15.3)	 & 386(14.1)    & -0.070\\
  \hspace{0.3cm} Never married & 378(27.7)	 & 801(29.3)    & 0.110\\
    \hspace{0.3cm} Missing & 12(0.9)	 & 10(0.4)    & 0.050\\
  \textbf{Education} \\
  \hspace{0.3cm} 0-8yrs & 124(9.1) & 278(10.2)  & -0.040\\ 
  \hspace{0.3cm} 9-11yrs & 187(13.7) & 359(13.2)  & 0.000\\ 
  \hspace{0.3cm} 12yrs & 525(38.5) & 1128(41.3) & 0.000\\ 
  \hspace{0.3cm} 13-15yrs & 216(15.8) & 422(15.5) & 0.010\\ 
  \hspace{0.3cm} $\geq 16$yrs & 172(12.6) & 291(10.7)  & 0.020\\ 
   \hspace{0.3cm} Missing & 141(10.3) & 252(9.2)  & 0.000\\
    \textbf{Annual family income} \\
  \hspace{0.3cm} $< \$1,000 $ & 38(2.8) & 54(2.0)  & 0.040\\ 
  \hspace{0.3cm} $ \$1,000$-$\$1,999 $ & 9(0.7) & 13(0.5)  & 0.020\\ 
  \hspace{0.3cm} $ \$2,000$-$\$2,999 $ & 8(0.6) & 20(0.7) &  -0.010\\ 
  \hspace{0.3cm} $ \$3,000$-$\$3,999 $ & 12(0.9) & 33(1.2) & -0.010\\ 
   \hspace{0.3cm} $ \$4,000$-$\$4,999 $ & 17(1.3) & 33(1.2) & -0.020\\ 
    \hspace{0.3cm} $ \$5,000$-$\$5,999 $ & 22(1.6) & 41(1.5) & -0.040\\ 
     \hspace{0.3cm} $ \$6,000$-$\$6,999 $ & 23(1.7) & 46(1.7) & -0.020\\ 
      \hspace{0.3cm} $ \$7,000$-$\$8,999 $ & 60(4.4) & 94(3.4) & 0.020\\ 
       \hspace{0.3cm} $ \$9,000$-$\$13,999 $ & 136(10.0) & 261(9.6) & 0.030\\ 
        \hspace{0.3cm} $ \$14,000$-$\$18,999 $ & 111(8.1) & 234(8.6) & -0.010\\ 
         \hspace{0.3cm} $ \$19,000$-$\$24,999 $ & 110(8.1) & 226(8.3) & -0.020\\ 
          \hspace{0.3cm} $ \$25,000$-$\$49,999 $ & 231(16.9) & 506(18.5) & -0.020\\ 
           \textbf{Veteran} & 355(26.0) & 518(19.0) &  0.120\\
   \textbf{Region} \\
  \hspace{0.3cm} Northeast & 157(11.5) & 345(12.6)  & -0.070\\ 
  \hspace{0.3cm} Midwest & 322(23.6)	 & 596(21.9)   & 0.050\\ 
  \hspace{0.3cm} South & 528(38.7) & 1114(40.8)  & -0.050\\ 
  \hspace{0.3cm} West & 358(26.2) & 675(24.7) & 0.060\\ 
   \textbf{Population} \\
  \hspace{0.3cm} Non-MSA & 367(26.9) & 763(28.0)  & -0.060\\ 
  \hspace{0.3cm} $<100,000$ & 12(0.8)	 & 22(0.8)   & 0.030\\ 
  \hspace{0.3cm} $100,000$-$249,999$ & 109(8.0) & 219(8.0)  & -0.040\\ 
  \hspace{0.3cm} $>250,000$ & 877(64.3) & 1725(63.2)  & 0.070\\ 
   \hline
\end{tabular}}
\end{table}

In the NMFS 1993 dataset, a total of 1,365 suicide deaths aged 18 years and older are identified. These suicide deaths are considered to be the narrow case, and are matched with 2,730 accidental deaths as referents ($1$ narrow case matched with $2$ marginal cases).  The matching process is designed to control over various variables, including alcohol usage status, sex, race, age, living arrangement, marital status, education, annual family income, military veteran status, geographical region, and population of the locality where the subject resided.  The matching first stratifies on sex and race to ensure exact match on these two variables and then minimizes a robust Mahalanobis distance; see \citet[Chapter 8]{rosenbaum2010design} for discussion of the multivariate matching. The balance and demographic distribution of the matched data are presented in Table \ref{tab:demographics}. We also apply the sample-splitting classification permutation test by \citet{chen2023testing}, and the result indicates that the matching is satisfactory. The Table \ref{tab:2x2 example} summarizes subjects are separated based on the status of having violent behavior with two types of cases (i.e., suicide death and accidental death). The odds ratio is 3.07 with $P$-value $< 2.2 \times 10^{-16}$. In a randomization test of no effect on the narrow cases, the one-sided $P$-value is $5.55 \times 10^{-16} $ with the attributable effect $A \geq 147$, which shows strong evidence that having violent behavior in the last year of life increases suicide risk. A sensitivity analysis shows that if there is no selection bias and the assumption of treatment does not cause marginal cases holds, then this causal conclusion is robust to hidden bias up to $\Gamma = 2.60$. $A \geq 147$ indicates that it is plausible that at least $147$ of the units having violent behavior in the last year of life completingsuicide are actually caused by having violent behavior in the last year of life. In the presence of selection bias, when $\Theta = 1.2$, it implies that up to $1 - 1/1.2 = 16.7\%$ of accidental deaths may transition to suicide deaths under the exposure of having violent behavior in the last year of life. A sensitivity analysis showed in the Table \ref{tab:ex} indicates that the observed effect is insensitive to bias up to $\Gamma = 2.60$, $\Theta = 1.00$ and $\Delta = 1.00$ or $\Gamma = 1.0$, $\Theta = 2.20$ and $\Delta = 1.20$ or other combinations.

\begin{table}
\caption{ The $2 \times 2$ contingency table of the study of having violent behavior in the last year of life and suicide risk, the odds ratio is $3.07$.}
\centering
\begin{tabular}{|| c c c c||}
\hline
  & Having violent behavior & Not having violent behavior & Total\\
 \hline\hline
 Suicide death & 247 & 1118 & 1365\\ 
 Accidental death & 183 & 2547 & 2730\\  
 Total & 430 & 3665  & 4095   \\
 \hline
\end{tabular}
\label{tab:2x2 example}
\end{table}

\begin{table}[!htbp]
\centering
\begin{tabular}{c c c c } 
 \hline
 $\Gamma$ & $\Theta$ &  $\Delta$ & $95\%$ minimum prediction interval   \\ 
 \hline
 $1.00$ & $1.00$ & 1.00  & $A \geq 147  $  \\ 
 $1.50$ & $1.00$ & 1.00 &$A \geq 101 $ \\ 
 $2.00$ & $1.00$ & 1.00 &$A \geq 54 $ \\ 
 $2.50$ & $1.00$ & 1.00 &$A \geq 8 $ \\ 
$\mathbf{2.60}$ & $\mathbf{1.00}$ &$\mathbf{1.00} $ &$A \mathbf{\geq 0 } $ \\ 
 $1.00$ & $1.20$ & 1.20 &$A \geq 106 $ \\ 
 $1.00$ & $1.40$ & 1.20 &$A \geq 84 $ \\ 
 $1.00$ & $1.60$ & 1.20 &$A \geq 61 $ \\ 
 $\mathbf{1.00}$ & $\mathbf{2.20}$ &$\mathbf{1.20} $ &$A \mathbf{\geq 0 } $ \\ 
 \hline
\end{tabular}
\caption{Various sensitivity analyses for statistical inference on the attributable effect in the study of having violent behavior in the last year of life and suicide risk. The last column report the $95\%$ minimum prediction interval of $A$ for different combinations of $\Gamma$, $\Theta$ and $\Gamma$. The bold number of $\Gamma$,$\Theta$ and $\Delta$ represents the maximum magnitude of sensitivity parameters that the rejection of the null hypothesis.  }
\label{tab:ex}
\end{table}

\subsection{Bounding Sensitivity Parameters}

We now provide an informal approach of bounding sensitivity parameter $\Theta$ and $\Delta$ using the 1993 NMFS data. Recall that in the sensitivity model \eqref{sensitivity model}, $\Theta$ and $\Delta$ are upper bounds of 
\begin{align*}
    \frac{\theta_{T_{ij}}}{\theta_{C_{ij}}}= \frac{\mathbb{P}(\kappa_n(R_{ij})=1\mid  x_{ij}, u_{ij}, r_{T_{ij}} =  r_{C_{ij}}= 1,  Z_{ij} = 1)}{\mathbb{P}(\kappa_n(R_{ij})=1\mid  x_{ij}, u_{ij}, r_{T_{ij}}=  r_{C_{ij}}= 1, Z_{ij}=0 )} 
\end{align*}
and 
\begin{align*}
         \frac{1 - \theta_{C_{ij}}}{1 - \theta_{T_{ij}}} = \frac{\mathbb{P}(\kappa_n(R_{ij})=0\mid  x_{ij}, u_{ij}, r_{T_{ij}}=  r_{C_{ij}}= 1,  Z_{ij} = 0)}{\mathbb{P}(\kappa_n(R_{ij})=0\mid  x_{ij}, u_{ij}, r_{T_{ij}}=  r_{C_{ij}}= 1, Z_{ij}=1 )},
\end{align*}
and recall that we assume the treatment has nonnegative effect (i.e., $ \kappa_n (r_{T_{ij}} ) \geq \kappa_n (r_{C_{ij}}) $ and $r_{T_{ij}} \geq r_{C_{ij}}$). In our data application, this means that having violent behavior in the last year of life will never prevent suicide deaths. Additionally, we also assume that 
\begin{align*}
    \mathbb{P}(\kappa_n(r_{T_{ij}}) | x_{ij},u_{ij}, r_{T_{ij}} = 1, r_{C_{ij}} = 0 ) \geq  \mathbb{P}(\kappa_n(r_{T_{ij}}) | x_{ij},u_{ij}, r_{T_{ij}} = r_{C_{ij}} = 1 )
\end{align*}
where the left-hand side term is the probability of completingsuicide if having violent behavior in the last year of life among units whose deaths (i.e., suicide deaths or accidental deaths) are attributable to having violent behavior in the last year of life. This left-hand side probability is likely to be very high because if a unit would only commit suicide if the unit had violent behavior in the last year of life, it is very likely that the unit would commit the suicide if he/she had violent behavior in the last year of life. Therefore, it is reasonable to assume that the left-hand side term is no less then the probability on the right-hand side, which is the probability of suicide if having violent behavior in the last year of life among always-cases. Then we could bound $\theta_{T_{ij}}/\theta_{C_{ij}}$ and $(1 - \theta_{C_{ij}})/(1 - \theta_{T_{ij}})$ using 
\begin{equation} \label{model: calibrate}
    \begin{split}
        &\frac{\mathbb{P}(\kappa_n(R_{ij})=1\mid  x_{ij}, u_{ij}, R_{ij}= 1,  Z_{ij} = 1)}{\mathbb{P}(\kappa_n(R_{ij})=1\mid  x_{ij}, u_{ij},R_{ij}= 1, Z_{ij}=0 )} \geq  \frac{\theta_{T_{ij}}}{\theta_{C_{ij}}}, \\
                &\frac{\mathbb{P}(\kappa_n(R_{ij})=0\mid  x_{ij}, u_{ij}, R_{ij}= 1,  Z_{ij} = 0)}{\mathbb{P}(\kappa_n(R_{ij})=0\mid  x_{ij}, u_{ij},R_{ij}= 1, Z_{ij}=1 )} \geq   \frac{1 - \theta_{C_{ij}}}{1 - \theta_{T_{ij}}}.
    \end{split}
\end{equation}
Now, we can calibrate the sensitivity parameter $\Theta$ and $\Delta$ by fitting a mixed effect logistic regression model to the left-hand side terms of \eqref{model: calibrate}. Specifically, we fit a mixed effect logistic regression model to $\kappa_n(R_{ij})$ on observed covariate $x_{ij}$, latent variable $u_{ij}$ and treatment assignment $Z_{ij}$. Then we used the fitted value to obtain upper bounds of these two ratios. 

In our data application, we found the maximum of these two ratios are 2.25 and 2.34 among all subjects which gives us an intuitive understanding of $\Theta$ and $\Delta$. It is important to acknowledge that this approach has limitations. First, $u_{ij}$ is a latent variable rather than unmeasured confounder when using mixed effect model approach because $u_{ij}$ needs to be independent of $x_{ij}$ and $Z_{ij}$. Second, the obtained values of $\Theta$ and $\Delta$ have estimation errors that were not accounting for. Formal approaches to calibrate the sensitivity parameter $\Theta$ and $\Delta$ can be viewed as a potential future research direction.

\section{Summary and Discussion}
\label{sec: conclusion}

In this article, we present a novel sensitivity analysis framework designed to address selection bias in matched case$^2$ studies, aiming to enhance the precision and reliability of effect attribution. Applied specifically to assess the attributable effect of violent behavior and firearms on suicide risk, our methodology systematically examines the potential impact of selection bias and unmeasured confounding bias. Additionally, the framework incorporates analysis for hidden bias, providing a comprehensive examination of potential sources of bias. Our contribution not only advances methodological approaches in case$^2$ studies but also holds promise for refining our understanding of complex associations in public health research, offering valuable insights to the broader field. For instance, in the investigation of the effect of alcohol abuse on breast cancer, the impact of alcohol consumption on hormone levels, such as estrogen, becomes a focal point. Notably, hormone-sensitive breast cancer cells possess receptors that utilize hormones for growth, suggesting that heightened hormone levels from alcohol consumption may elevate the likelihood of hormone-sensitive invasive breast cancer \citep{liu2015links}. Supporting this, \citet{li2010alcohol} present evidence from the Women's Health Initiative Observational Study, indicating a stronger association between alcohol consumption and hormone-sensitive breast cancer compared to hormone-insensitive breast cancer. Consequently, there arises the possibility that alcohol consumption may alter hormone-insensitive breast cancer to hormone-sensitive breast cancer, underscoring the imperative for the development of sensitivity analysis methodologies to address potential selection biases \citep{rosenberg1990case}. Therefore, we could leverage our framework for the investigation of the effect of alcohol consumption on the hormone-sensitive breast cancer in comparison to hormone-insensitive breast cancer.

%In our future work, we aim to extend the applicability of our proposed sensitivity analysis framework concerning selection bias to matched cohort studies. The inherent principles and methodologies encapsulated in our framework, originally tailored for matched case$^2$ studies, demonstrate versatility in effectively addressing biases within the context of matched observational designs. In addition, the development of a formal calibration for the sensitivity parameter could also be viewed as a future work. 

\newpage

\medskip

 \bibliographystyle{apalike}

\bibliography{reference}

\newpage

\begin{center}
    \Large Supplementary Materials for ``Sensitivity Analysis for Attributable Effects in Case$^2$ Studies"
\end{center}

\section*{Proof of Proposition \ref{prop: bound}}

\begin{proof}
From the sensitivity model (\ref{sensitivity model}), we have the following:

\begin{itemize}
    \item[(1)] $\mathbb{P} (Z_{ij} = 1, \kappa_n (R_{ij}) = 0 | R_{ij} = 1, x_{ij}, u_{ij}  ) = \pi_{ij} (1 - \theta_{T_{ij}}) $ 
    \item[(2)] $\mathbb{P} (\kappa_n (R_{ij}) = 1 | R_{ij} = 1, x_{ij}, u_{ij}  ) = \pi_{ij} \theta_{T_{ij}} + (1 - \pi_{ij}) \theta_{C_{ij}}$
\end{itemize}

We first calculate $\mathbb{P}(Z_{ij} = 1 | \mathcal{F}, \mathcal{B} )$ for $j = 1, 2, \cdots, J$.

\begin{align*}
    \mathbb{P}(Z_{ij} = 1 | \mathcal{F}, \mathcal{B} ) &= \mathcal{P} (Z_{ij} = 1| R_{ij} = 1, \kappa_{n}(R_{ij}) = 0, x_{ij}, u_{ij} ) \\ 
    &= \frac{\mathbb{P} (Z_{ij} = 1, \kappa_n (R_{ij}) = 0 | R_{ij} = 1, x_{ij}, u_{ij}  )   }{\mathbb{P} (\kappa_n (R_{ij}) = 0 | R_{ij} = 1, x_{ij}, u_{ij}  )} \\
    & = \frac{\pi_{ij} (1 - \theta_{T_{ij}})  }{1 - \pi_{ij} \theta_{T_{ij}} - (1 - \pi_{ij}) \theta_{C_{ij}} }
\end{align*}
for $j = 2, \cdots, J$. And
\begin{align*}
     \mathbb{P}(Z_{i1} = 1 | \mathcal{F}, \mathcal{B} ) &= \mathcal{P} (Z_{i1} = 1| R_{i1} = \kappa_{n}(R_{i1}) = 1, x_{i1}, u_{i1} ) \\ 
    &= \frac{\mathbb{P} (Z_{i1} = 1, \kappa_n (R_{i1}) = 1 | R_{i1} = 1, x_{i1}, u_{i1}  )   }{\mathbb{P} (\kappa_n (R_{i1}) = 1 | R_{i1} = 1, x_{i1}, u_{i1}  )} \\
    & = \frac{\pi_{i1} \theta_{T_{i1}}  }{\pi_{i1} \theta_{T_{i1}} + (1 - \pi_{i1}) \theta_{C_{i1}} }.
\end{align*}
for $j = 1$. If $Z_{i^+} = 0, \mathbb{P}(C_i = 1| \mathcal{F},\mathcal{B}, \mathcal{Z}) = 0$; if $Z_{i^+} = J, \mathbb{P}(C_i = 1| \mathcal{F},\mathcal{B}, \mathcal{Z}) = 1$. For $1 \leq Z_{i^+} \leq J-1$, let $\mathcal{G}(J,Z_{i^+})$ be the set containing the $\binom{J}{Z_{i^+}}$ possible values of $\mathbf{Z}_i$, then

\begin{align*}
    &\mathbb{P}(C_i = 1| \mathcal{F},\mathcal{B}, \mathcal{Z})  \\
    & = \mathbb{P}(Z_{i1} = 1| \mathcal{F},\mathcal{B}, \mathcal{Z}) \\
    & = \frac{\mathbb{P}(Z_{i1} = 1, \mathbf{Z}_i \in \mathcal{Z}| \mathcal{F},\mathcal{B} ) }{\mathbb{P}(\mathbf{Z}_i \in \mathcal{Z}| \mathcal{F},\mathcal{B} ) } \\
    & = \frac{\mathbb{P}(Z_{i1} = 1, \mathbf{Z}_i \in \mathcal{Z}| \mathcal{F},\mathcal{B} ) }{\mathbb{P}(Z_{i1} = 1, \mathbf{Z}_i \in \mathcal{Z}| \mathcal{F},\mathcal{B} ) + \mathbb{P}(Z_{i1} = 0, \mathbf{Z}_i \in \mathcal{Z}| \mathcal{F},\mathcal{B} ) } \\
    & = \frac{\sum_{\mathbf{z} \in \mathcal{G}(J-1,Z_{i^+}-1) } \mathbb{P}(Z_{i1} = 1, \mathbf{Z}_{i,-1} = \mathbf{z}| \mathcal{F},\mathcal{B} ) }{\sum_{\mathbf{z} \in \mathcal{G}(J-1,Z_{i^+}-1) } \mathbb{P}(Z_{i1} = 1, \mathbf{Z}_{i,-1} = \mathbf{z}| \mathcal{F},\mathcal{B} ) + \sum_{\mathbf{z} \in \mathcal{G}(J-1,Z_{i^+}) } \mathbb{P}(Z_{i1} = 0, \mathbf{Z}_{i,-1} = \mathbf{z}| \mathcal{F},\mathcal{B} ) } \\
    & = \frac{\sum_{\mathbf{z} \in \mathcal{G}(J-1,Z_{i^+}-1) } \mathbb{P}(Z_{i1} = 1| \mathcal{F},\mathcal{B} ) \mathbb{P}(\mathbf{Z}_{i,-1} = \mathbf{z}| \mathcal{F},\mathcal{B} ) }{\sum_{\mathbf{z} \in \mathcal{G}(J-1,Z_{i^+}-1) } \mathbb{P}(Z_{i1} = 1| \mathcal{F},\mathcal{B} ) \mathbb{P}(\mathbf{Z}_{i,-1} = \mathbf{z}| \mathcal{F},\mathcal{B} ) + \sum_{\mathbf{z} \in \mathcal{G}(J-1,Z_{i^+}) } \mathbb{P}(Z_{i1} = 0| \mathcal{F},\mathcal{B} ) \mathbb{P}(\mathbf{Z}_{i,-1} = \mathbf{z}| \mathcal{F},\mathcal{B} ) } \\
    & = \frac{\mathbb{P}(Z_{i1} = 1| \mathcal{F},\mathcal{B} ) \sum_{\mathbf{z} \in \mathcal{G}(J-1,Z_{i^+}-1) } \mathbb{P}(\mathbf{Z}_{i,-1} = \mathbf{z}| \mathcal{F},\mathcal{B} ) }{\mathbb{P}(Z_{i1} = 1| \mathcal{F},\mathcal{B} )\sum_{\mathbf{z} \in \mathcal{G}(J-1,Z_{i^+}-1) }  \mathbb{P}(\mathbf{Z}_{i,-1} = \mathbf{z}| \mathcal{F},\mathcal{B} ) + \mathbb{P}(Z_{i1} = 0| \mathcal{F},\mathcal{B} )\sum_{\mathbf{z} \in \mathcal{G}(J-1,Z_{i^+}) }  \mathbb{P}(\mathbf{Z}_{i,-1} = \mathbf{z}| \mathcal{F},\mathcal{B} ) } \\
    & = \frac{\pi_{i1} \theta_{T_{i1}} \sum_{\mathbf{z} \in \mathcal{G}(J-1,Z_{i^+}-1) } \mathbb{P}(\mathbf{Z}_{i,-1} = \mathbf{z}| \mathcal{F},\mathcal{B} ) }{\pi_{i1} \theta_{T_{i1}}\sum_{\mathbf{z} \in \mathcal{G}(J-1,Z_{i^+}-1) }  \mathbb{P}(\mathbf{Z}_{i,-1} = \mathbf{z}| \mathcal{F},\mathcal{B} ) + (1 - \pi_{i1})\theta_{C_{i1}}\sum_{\mathbf{z} \in \mathcal{G}(J-1,Z_{i^+}) }  \mathbb{P}(\mathbf{Z}_{i,-1} = \mathbf{z}| \mathcal{F},\mathcal{B} ) } \\
    & =  \frac{\pi_{i1} \theta_{T_{i1}} \sum_{\mathbf{z} \in \mathcal{G}(J-1,Z_{i^+}-1) } \prod_{j=2}^J \left( \frac{\pi_{ij} (1 - \theta_{T_{ij}})  }{1 - \pi_{ij} \theta_{T_{ij}} - (1 - \pi_{ij}) \theta_{C_{ij}} } \right)^{z_{ij}} \left(1 - \frac{\pi_{ij} (1 - \theta_{T_{ij}})  }{1 - \pi_{ij} \theta_{T_{ij}} - (1 - \pi_{ij}) \theta_{C_{ij}} } \right)^{1 - z_{ij}} }{\splitfrac{\pi_{i1} \theta_{T_{i1}}\sum_{\mathbf{z} \in \mathcal{G}(J-1,Z_{i^+}-1) }  \prod_{j=2}^J \left( \frac{\pi_{ij} (1 - \theta_{T_{ij}})  }{1 - \pi_{ij} \theta_{T_{ij}} - (1 - \pi_{ij}) \theta_{C_{ij}} } \right)^{z_{ij}} \left(1 - \frac{\pi_{ij} (1 - \theta_{T_{ij}})  }{1 - \pi_{ij} \theta_{T_{ij}} - (1 - \pi_{ij}) \theta_{C_{ij}} } \right)^{1 - z_{ij}} } {+(1 - \pi_{i1})\theta_{C_{i1}}\sum_{\mathbf{z} \in \mathcal{G}(J-1,Z_{i^+}) }  \prod_{j=2}^J \left( \frac{\pi_{ij} (1 - \theta_{T_{ij}})  }{1 - \pi_{ij} \theta_{T_{ij}} - (1 - \pi_{ij}) \theta_{C_{ij}} } \right)^{z_{ij}} \left(1 - \frac{\pi_{ij} (1 - \theta_{T_{ij}})  }{1 - \pi_{ij} \theta_{T_{ij}} - (1 - \pi_{ij}) \theta_{C_{ij}} } \right)^{1 - z_{ij}} }}.  
\end{align*}
Note that
\begin{align*}
    & \qquad\frac{\pi_{ij} (1 - \theta_{T_{ij}})  }{1 - \pi_{ij} \theta_{T_{ij}} - (1 - \pi_{ij}) \theta_{C_{ij}} }  \\
    &= \frac{\pi_{ij} (1 - \theta_{T_{ij}}) }{(1 - \theta_{C_{ij}}) - \pi_{ij} \left((1 - \theta_{C_{ij}}) - (1 - \theta_{T_{ij}})  \right) } \\
    & = \frac{\pi_{ij}}{\frac{1 - \theta_{C_{ij}}}{1 - \theta_{T_{ij}}} - \pi_{ij}(\frac{1 - \theta_{C_{ij}}}{1 - \theta_{T_{ij}}} - 1 )   } = \frac{\pi_{ij}}{\Omega_{ij} - \pi_{ij} (\Omega_{ij} - 1) }
\end{align*}
where we denote $\frac{1 - \theta_{C_{ij}}}{1 - \theta_{T_{ij}}} = \Omega_{ij}$. Then, 
\begin{align*}
  & \mathbb{P}(Z_{i1}=1 | \mathcal{F},\mathcal{B}, \mathcal{Z}) \\
    & =  \frac{\pi_{i1} \theta_{T_{i1}} \sum_{\mathbf{z} \in \mathcal{G}(J-1,Z_{i^+}-1) } \prod_{j=2}^J \left( \frac{\pi_{ij} (1 - \theta_{T_{ij}})  }{1 - \pi_{ij} \theta_{T_{ij}} - (1 - \pi_{ij}) \theta_{C_{ij}} } \right)^{z_{ij}} \left(1 - \frac{\pi_{ij} (1 - \theta_{T_{ij}})  }{1 - \pi_{ij} \theta_{T_{ij}} - (1 - \pi_{ij}) \theta_{C_{ij}} } \right)^{1 - z_{ij}} }{\splitfrac{\pi_{i1} \theta_{T_{i1}}\sum_{\mathbf{z} \in \mathcal{G}(J-1,Z_{i^+}-1) }  \prod_{j=2}^J \left( \frac{\pi_{ij} (1 - \theta_{T_{ij}})  }{1 - \pi_{ij} \theta_{T_{ij}} - (1 - \pi_{ij}) \theta_{C_{ij}} } \right)^{z_{ij}} \left(1 - \frac{\pi_{ij} (1 - \theta_{T_{ij}})  }{1 - \pi_{ij} \theta_{T_{ij}} - (1 - \pi_{ij}) \theta_{C_{ij}} } \right)^{1 - z_{ij}} } {+(1 - \pi_{i1})\theta_{C_{i1}}\sum_{\mathbf{z} \in \mathcal{G}(J-1,Z_{i^+}) }  \prod_{j=2}^J \left( \frac{\pi_{ij} (1 - \theta_{T_{ij}})  }{1 - \pi_{ij} \theta_{T_{ij}} - (1 - \pi_{ij}) \theta_{C_{ij}} } \right)^{z_{ij}} \left(1 - \frac{\pi_{ij} (1 - \theta_{T_{ij}})  }{1 - \pi_{ij} \theta_{T_{ij}} - (1 - \pi_{ij}) \theta_{C_{ij}} } \right)^{1 - z_{ij}} }} \\
    & = \frac{\pi_{i1} \theta_{T_{i1}} \sum_{\mathbf{z} \in \mathcal{G}(J-1,Z_{i^+}-1) } \left( \prod_{j=2}^J \pi_{ij}^{z_{ij}} (1 - \pi_{ij})^{1-z_{ij}} \right) \left( \prod_{j=2}^J \frac{\Omega_{ij}^{1-z_{ij}}}{\Omega_{ij} - \pi_{ij}(\Omega_{ij} - 1) } \right) }{\splitfrac{\pi_{i1} \theta_{T_{i1}}\sum_{\mathbf{z} \in \mathcal{G}(J-1,Z_{i^+}-1) }  \left( \prod_{j=2}^J \pi_{ij}^{z_{ij}} (1 - \pi_{ij})^{1-z_{ij}} \right) \left( \prod_{j=2}^J \frac{\Omega_{ij}^{1-z_{ij}}}{\Omega_{ij} - \pi_{ij}(\Omega_{ij} - 1) } \right) } {+(1 - \pi_{i1})\theta_{C_{i1}}\sum_{\mathbf{z} \in \mathcal{G}(J-1,Z_{i^+}) }  \left( \prod_{j=2}^J \pi_{ij}^{z_{ij}} (1 - \pi_{ij})^{1-z_{ij}} \right) \left( \prod_{j=2}^J \frac{\Omega_{ij}^{1-z_{ij}}}{\Omega_{ij} - \pi_{ij}(\Omega_{ij} - 1) } \right) }} \\
    & = \frac{\pi_{i1} \theta_{T_{i1}} \sum_{\mathbf{z} \in \mathcal{G}(J-1,Z_{i^+}-1) } \left( \prod_{j=2}^J \pi_{ij}^{z_{ij}} (1 - \pi_{ij})^{1-z_{ij}} \right) \left( \prod_{j=2}^J \Omega_{ij}^{1-z_{ij}} \right) \left( \prod_{j=2}^J  \frac{1}{\Omega_{ij} - \pi_{ij}(\Omega_{ij} - 1)}  \right) }{\splitfrac{\pi_{i1} \theta_{T_{i1}}\sum_{\mathbf{z} \in \mathcal{G}(J-1,Z_{i^+}-1) }  \left( \prod_{j=2}^J \pi_{ij}^{z_{ij}} (1 - \pi_{ij})^{1-z_{ij}} \right) \left( \prod_{j=2}^J \Omega_{ij}^{1-z_{ij}} \right) \left( \prod_{j=2}^J  \frac{1}{\Omega_{ij} - \pi_{ij}(\Omega_{ij} - 1)}  \right) } {+(1 - \pi_{i1})\theta_{C_{i1}}\sum_{\mathbf{z} \in \mathcal{G}(J-1,Z_{i^+}) }  \left( \prod_{j=2}^J \pi_{ij}^{z_{ij}} (1 - \pi_{ij})^{1-z_{ij}} \right) \left( \prod_{j=2}^J \Omega_{ij}^{1-z_{ij}} \right) \left( \prod_{j=2}^J  \frac{1}{\Omega_{ij} - \pi_{ij}(\Omega_{ij} - 1)}  \right) }} \\
    & = \frac{\pi_{i1} \theta_{T_{i1}} \sum_{\mathbf{z} \in \mathcal{G}(J-1,Z_{i^+}-1) } \left( \prod_{j=2}^J \pi_{ij}^{z_{ij}} (1 - \pi_{ij})^{1-z_{ij}} \right) \left( \prod_{j=2}^J \Omega_{ij}^{1-z_{ij}} \right)  }{\splitfrac{\pi_{i1} \theta_{T_{i1}}\sum_{\mathbf{z} \in \mathcal{G}(J-1,Z_{i^+}-1) }  \left( \prod_{j=2}^J \pi_{ij}^{z_{ij}} (1 - \pi_{ij})^{1-z_{ij}} \right) \left( \prod_{j=2}^J \Omega_{ij}^{1-z_{ij}} \right) } {+(1 - \pi_{i1})\theta_{C_{i1}}\sum_{\mathbf{z} \in \mathcal{G}(J-1,Z_{i^+}) }  \left( \prod_{j=2}^J \pi_{ij}^{z_{ij}} (1 - \pi_{ij})^{1-z_{ij}} \right) \left( \prod_{j=2}^J \Omega_{ij}^{1-z_{ij}} \right)  }}  \\
    & = \frac{\pi_{i1} \theta_{T_{i1}} \sum_{\mathbf{z} \in \mathcal{G}(J-1,Z_{i^+}-1) } \left( \prod_{j=2}^J \pi_{ij}^{z_{ij}} (1 - \pi_{ij})^{1-z_{ij}} \right) \Omega_{ij}^{J-Z_{i^+}}  }{\splitfrac{\pi_{i1} \theta_{T_{i1}}\sum_{\mathbf{z} \in \mathcal{G}(J-1,Z_{i^+}-1) }  \left( \prod_{j=2}^J \pi_{ij}^{z_{ij}} (1 - \pi_{ij})^{1-z_{ij}} \right) \Omega_{ij}^{J-Z_{i^+}} } {+(1 - \pi_{i1})\theta_{C_{i1}}\sum_{\mathbf{z} \in \mathcal{G}(J-1,Z_{i^+}) }  \left( \prod_{j=2}^J \pi_{ij}^{z_{ij}} (1 - \pi_{ij})^{1-z_{ij}} \right) \Omega_{ij}^{J-Z_{i^+} - 1}  }} \\
    & = \frac{\pi_{i1} \theta_{T_{i1}} \sum_{\mathbf{z} \in \mathcal{G}(J-1,Z_{i^+}-1) } \left( \prod_{j=2}^J \pi_{ij}^{z_{ij}} (1 - \pi_{ij})^{1-z_{ij}} \right)  }{\splitfrac{\pi_{i1} \theta_{T_{i1}}\sum_{\mathbf{z} \in \mathcal{G}(J-1,Z_{i^+}-1) }  \left( \prod_{j=2}^J \pi_{ij}^{z_{ij}} (1 - \pi_{ij})^{1-z_{ij}} \right)  } {+(1 - \pi_{i1})\theta_{C_{i1}}\sum_{\mathbf{z} \in \mathcal{G}(J-1,Z_{i^+}) }  \left( \prod_{j=2}^J \pi_{ij}^{z_{ij}} (1 - \pi_{ij})^{1-z_{ij}} \right) \Omega_{ij}^{-1}  }} \\
    & \leq \frac{\pi_{i1} \theta_{T_{i1}} \sum_{\mathbf{z} \in \mathcal{G}(J-1,Z_{i^+}-1) } \left( \prod_{j=2}^J \pi_{ij}^{z_{ij}} (1 - \pi_{ij})^{1-z_{ij}} \right)  }{\splitfrac{\pi_{i1} \theta_{T_{i1}}\sum_{\mathbf{z} \in \mathcal{G}(J-1,Z_{i^+}-1) }  \left( \prod_{j=2}^J \pi_{ij}^{z_{ij}} (1 - \pi_{ij})^{1-z_{ij}} \right)  } {+(1 - \pi_{i1})\theta_{C_{i1}}\sum_{\mathbf{z} \in \mathcal{G}(J-1,Z_{i^+}) }  \left( \prod_{j=2}^J \pi_{ij}^{z_{ij}} (1 - \pi_{ij})^{1-z_{ij}} \right) \frac{1}{\Delta }  }} \\
    & = \frac{ \theta_{T_{i1}} \exp (\log (\Gamma)u_{i1} ) \sum_{\mathbf{z} \in \mathcal{G}(J-1,Z_{i^+}-1) }\exp (\log (\Gamma)\sum_{j=2}^J z_{ij} u_{ij} )  }{\splitfrac{\theta_{T_{i1}} \exp (\log (\Gamma)u_{i1} ) \sum_{\mathbf{z} \in \mathcal{G}(J-1,Z_{i^+}-1) }\exp (\log (\Gamma)\sum_{j=2}^J z_{ij} u_{ij} )   } {+(1 - \pi_{i1})\theta_{C_{i1}}\sum_{\mathbf{z} \in \mathcal{G}(J-1,Z_{i^+}) }  \exp (\log (\Gamma)\sum_{j=2}^J z_{ij} u_{ij} )  \frac{1}{\Delta }  }}
\end{align*}
since $\frac{1}{\Delta } \leq \frac{1 - \theta_{T_{ij}}}{1 - \theta_{C_{ij}}}$ by sensitivity analysis model. And 

\begin{align*}
  &\qquad    \mathbb{P}(Z_{i1}=1 | \mathcal{F},\mathcal{B}, \mathcal{Z}) \\
      & = \frac{\pi_{i1} \theta_{T_{i1}} \sum_{\mathbf{z} \in \mathcal{G}(J-1,Z_{i^+}-1) } \left( \prod_{j=2}^J \pi_{ij}^{z_{ij}} (1 - \pi_{ij})^{1-z_{ij}} \right)  }{\splitfrac{\pi_{i1} \theta_{T_{i1}}\sum_{\mathbf{z} \in \mathcal{G}(J-1,Z_{i^+}-1) }  \left( \prod_{j=2}^J \pi_{ij}^{z_{ij}} (1 - \pi_{ij})^{1-z_{ij}} \right)  } {+(1 - \pi_{i1})\theta_{C_{i1}}\sum_{\mathbf{z} \in \mathcal{G}(J-1,Z_{i^+}) }  \left( \prod_{j=2}^J \pi_{ij}^{z_{ij}} (1 - \pi_{ij})^{1-z_{ij}} \right) \Omega_{ij}^{-1}  }} \\
      & \geq \frac{\pi_{i1} \theta_{T_{i1}} \sum_{\mathbf{z} \in \mathcal{G}(J-1,Z_{i^+}-1) } \left( \prod_{j=2}^J \pi_{ij}^{z_{ij}} (1 - \pi_{ij})^{1-z_{ij}} \right)  }{\splitfrac{\pi_{i1} \theta_{T_{i1}}\sum_{\mathbf{z} \in \mathcal{G}(J-1,Z_{i^+}-1) }  \left( \prod_{j=2}^J \pi_{ij}^{z_{ij}} (1 - \pi_{ij})^{1-z_{ij}} \right)  } {+(1 - \pi_{i1})\theta_{C_{i1}}\sum_{\mathbf{z} \in \mathcal{G}(J-1,Z_{i^+}) }  \left( \prod_{j=2}^J \pi_{ij}^{z_{ij}} (1 - \pi_{ij})^{1-z_{ij}} \right)  }} \\
      & =  \frac{ \theta_{T_{i1}} \exp (\log (\Gamma)u_{i1} ) \sum_{\mathbf{z} \in \mathcal{G}(J-1,Z_{i^+}-1) }\exp (\log (\Gamma)\sum_{j=2}^J z_{ij} u_{ij} )  }{\splitfrac{\theta_{T_{i1}} \exp (\log (\Gamma)u_{i1} ) \sum_{\mathbf{z} \in \mathcal{G}(J-1,Z_{i^+}-1) }\exp (\log (\Gamma)\sum_{j=2}^J z_{ij} u_{ij} )   } {+(1 - \pi_{i1})\theta_{C_{i1}}\sum_{\mathbf{z} \in \mathcal{G}(J-1,Z_{i^+}) }  \exp (\log (\Gamma)\sum_{j=2}^J z_{ij} u_{ij} ) }}.
\end{align*}

because $\frac{1 - \theta_{T_{ij}}}{1 - \theta_{C_{ij}}} \leq 1$ by sensitivity analysis model again. Write $w_{ij} = \exp (\log(\Gamma)u_{ij} ) > 0$ and $\bm w_i = (w_{i2}, \cdots, w_{iJ})^\top$. Then we have
\begin{align*}
    \sum_{\mathbf{z} \in \mathcal{G}(J-1,b) }  \exp \left(\log (\Gamma)\sum_{j=2}^J z_{ij} u_{ij} \right) = \sum_{\mathbf{z} \in \mathcal{G}(J-1,b) }  \prod_{j=2}^J w_{ij}^{z_{ij}} := S_b (\bm w_i).
\end{align*}

where $S_b (\bm w_i)$ is the $b$th elementary symmetric functions of $\bm w_i$ .Using $S_b (\bm w_i)$, we have
\begin{align*}
    \mathbb{P} (C_i = 1 | \mathcal{F}, \mathcal{B}, \mathcal{Z}) &\leq \frac{\theta_{T_{i1}} \exp (\log (\Gamma)u_{i1} ) S_{Z_{i^+} - 1}(\bm w_i)  }{\theta_{T_{i1}} \exp (\log (\Gamma)u_{i1} ) S_{Z_{i^+} - 1}(\bm w_i) + \theta_{C_{i1}} S_{Z_{i^+}}(\bm w_i) \frac{1}{\Delta} }  \\
    & = \frac{ \frac{\theta_{T_{i1}}}{\theta_{C_{i1}}} \exp (\log (\Gamma)u_{i1} )   }{\frac{\theta_{T_{i1}}}{\theta_{C_{i1}}} \exp (\log (\Gamma)u_{i1} )  +   \frac{S_{Z_{i^+}}(\bm w_i)}{S_{Z_{i^+} - 1}(\bm w_i)} \frac{1}{\Delta} } 
\end{align*}
and
\begin{align*}
    \mathbb{P} (C_i = 1 | \mathcal{F}, \mathcal{B}, \mathcal{Z}) &\geq \frac{\theta_{T_{i1}} \exp (\log (\Gamma)u_{i1} ) S_{Z_{i^+} - 1}(\bm w_i)  }{\theta_{T_{i1}} \exp (\log (\Gamma)u_{i1} ) S_{Z_{i^+} - 1}(\bm w_i) + \theta_{C_{i1}} S_{Z_{i^+}}(\bm w_i) }  \\
    & = \frac{\frac{\theta_{T_{i1}}}{\theta_{C_{i1}}} \exp (\log (\Gamma)u_{i1} )   }{\frac{\theta_{T_{i1}}}{\theta_{C_{i1}}} \exp (\log (\Gamma)u_{i1} )  +   \frac{S_{Z_{i^+}}(\bm w_i)}{S_{Z_{i^+} - 1}(\bm w_i)}  }.
\end{align*}
Note that $\mathbb{P} (C_i = 1 | \mathcal{F}, \mathcal{B}, \mathcal{Z})$ is monotone in $\Theta \exp (\log(\Gamma) u_{i1} )$. Also, from the property of elementary symmetric functions, $\frac{S_{Z_{i^+}}(\bm w_i)}{S_{Z_{i^+} - 1}(\bm w_i)} $ is increasing in each $w_{ij}$, and thus also in each $u_{ij}$, $2 \leq j \leq J$. So we have

\begin{align*}
  \bar{p}_{ni} =  \frac{Z_{i^+}}{Z_{i^+} + (J - Z_{i^+}) \Gamma } \leq \mathbb{P} (C_i = 1 | \mathcal{F}, \mathcal{B}, \mathcal{Z}) \leq \frac{Z_{i^+} \Theta \Delta \Gamma }{Z_{i^+} \Theta \Delta \Gamma + (J - Z_{i^+})} =\bar{\bar{p}}_{ni} 
\end{align*}
under the null hypothesis. The lower bound is attained at $\bm u_i = (0,1,\cdots,1) $ and $(1 - \theta_{C_{ij}})/(1 - \theta_{T_{ij}})  = \theta_{T_{ij}}/ \theta_{C_{ij}} = 1$, and the upper bound is attained at $\bm u_i = (1,0,\cdots,0) $, $ \theta_{T_{ij}}/ \theta_{C_{ij}} = \Theta$ and $(1 - \theta_{C_{ij}})/(1 - \theta_{T_{ij}}) = \Delta$. 
\end{proof}

\section*{Violation of Nonnegative Treatment Effect Assumption}

It is possible that some of the units may violate this non-negativity assumption. However, if we know the maximum number of units that violate this assumption, valid statistical inference can be still conducted. To see this, we first look at the $2 \times 2$ contingency table below. 

\begin{table}[!ht]
\label{tab: AE and CS}
\caption{ A $2 \times 2$ table adjusted for attributable effects }
\centering
\begin{tabular}{|| c c c c||}
\hline
 Response & Treated  & Control  & Total   \\
 \hline\hline
1 & $a-A$ & $b$  & $a+b-A$  \\ 
0 & $c$ & $d$ & $c+d$  \\  
Total & $a+c-A$ & $b+d$    & $N-A$  \\
 \hline
\end{tabular}
\end{table}

Suppose there are $n$ units violating the non-negativity assumption, then these $n$ units have to come from $b$ and $d$ because they are cases under control and all units in the table are cases. We want to subtract $n$ from the table and conduct inference on the attributable effect. Without loss of generality, assuming that $b > n$ and $d > n$, then there are $n+1$ possibilities of allocations such that $b_i$ units comes from $b$ units, $d_i$ units comes from $d$ units and $b_i + d_i = n$ where $i = 1,2,\cdots, n + 1$. Denote $m$ be the index of the allocation that generate the largest $P$-value among all $i$ using Fisher's exact test, i.e., 

\begin{align*}
    \frac{\binom{a+b-b_{m}}{a} \binom{c+d-d_{m}}{c}     }{\binom{N-b_{m}-d_{m}}{a+b-c_{m}} } = \max_{i \in \{1,2,\cdots,n+1 \}} \left\{  \frac{\binom{a+b-b_{i}}{a} \binom{c+d-d_{i}}{c}     }{\binom{N-b_{i}-d_{i}}{a+b-c_{i}} }     \right\}
\end{align*}
and

\begin{align*}
    \frac{\binom{a+b-b_{i^*}}{a} \binom{c+d-d_{i^*}}{c}     }{\binom{N-b_{i^*}-d_{i^*}}{a+b-c_{i^*}} }
\end{align*}
be the $P$-value of the ground-truth allocation. By construction, 
\begin{align*}
     \frac{\binom{a+b-b_{m}}{a} \binom{c+d-d_{m}}{c}     }{\binom{N-b_{m}-d_{m}}{a+b-c_{m}} } \geq  \frac{\binom{a+b-b_{i^*}}{a} \binom{c+d-d_{i^*}}{c}     }{\binom{N-b_{i^*}-d_{i^*}}{a+b-c_{i^*}} }.
\end{align*}
and hence
\begin{align*}
     \frac{\binom{a-A+b-b_{m}}{a-A} \binom{c+d-d_{m}}{c}     }{\binom{N-A-b_{m}-d_{m}}{a-A+b-c_{m}} } \geq  \frac{\binom{a-A+b-b_{i^*}}{a-A} \binom{c+d-d_{i^*}}{c}     }{\binom{N-A-b_{i^*}-d_{i^*}}{a-A+b-c_{i^*}} }.
\end{align*}
Therefore, the one-sided prediction interval for the attributable effect obtained from the $m$-th allocation will always cover the one-sided prediction interval for the attributable effect obtained from the $i^*$-th allocation which is the ground-truth allocation. Similar idea can be implemented by Mantel-Haenszel test as well across different matched sets. This can be viewed as a sensitivity analysis where the number of units that violate the non-negativity assumption is the sensitivity parameter. 

\newpage

\end{document}